\documentclass{mn2e}

\input epsf

\title[CLASS statistical lensing, cosmological parameters, and global 
properties of galaxy populations]
{The Cosmic Lens All-Sky Survey: statistical strong lensing, cosmological
parameters, and global properties of galaxy populations}

\author[Kyu-Hyun Chae]
{Kyu-Hyun Chae \\
Sejong University, Department of Astronomy and Space Sciences,
98 Gunja-dong, Gwangjin-Gu, Seoul 143-747, Republic of Korea;\\
University of Manchester, Jodrell Bank Observatory,
Macclesfield, Cheshire SK11 9DL, UK}

\date{Accepted ........
      Received .......;
      in original form .......}

\pubyear{2003}

\begin{document}
\maketitle
\begin{abstract}
Extensive analyses of statistical strong gravitational lensing are performed
based on the final Cosmic Lens All-Sky Survey (CLASS) well-defined statistical
sample of flat-spectrum radio sources and current estimates of galaxy
luminosity functions per morphological type. The analyses are done under
the assumption that galactic lenses are well-approximated by singular
isothermal ellipsoids and early-type galaxies evolved passively since
redshift $z \sim 1$. Two goals of the analyses are:
(1) to constrain cosmological parameters independently of other techniques 
(e.g.\ Type~Ia supernovae magnitude-redshift relation, 
cosmic microwave background anisotropies, galaxy matter power spectra); 
and (2) to constrain the characteristic line-of-sight
velocity dispersion and the mean projected mass ellipticity for the early-type
galaxy population. Depending on how the late-type galaxy population is
treated (i.e., whether its characteristic velocity dispersion is constrained
or not), we find for a flat universe with a classical cosmological constant
that the matter fraction of the present critical density
$\Omega_{\rm m} = 0.31^{+0.27}_{-0.14}$ (68\%) for the unconstrained
case or $0.40^{+0.28}_{-0.16}$ (68\%) for the constrained case,
with an additional systematic uncertainty of $\approx 0.11$ arising from
the present uncertainty in the distribution of CLASS sources in redshift and
flux density. For a flat universe with a constant equation of state for
dark energy $w$~=~$p_x$(pressure)/$\rho_x$(energy density) and the 
prior constraint $w \ge -1$, we find that
$-1 \le w < -0.55^{+0.18}_{-0.11}$ (68\%) for the unconstrained case or
$-1 \le w < -0.41^{+0.28}_{-0.16}$ (68\%) for the constrained case, where 
$w=-1$ corresponds to a classical cosmological constant.
The determined value of the early-type characteristic velocity dispersion
($\sigma_{*}^{(e)}$) depends on the faint-end slope of the early-type
luminosity function ($\alpha^{(e)}$) and the intrinsic shape distribution
of galaxies; for equal frequencies of oblates and prolates, we find that
$\sigma_{*}^{(e)}(0.3 \la z \la 1)= 198^{+22}_{-18}$~km~s$^{-1}$ (68\%)
for a `steep' $\alpha^{(e)}=-1$ or
$\sigma_{*}^{(e)}(0.3 \la z \la 1) = 181^{+18}_{-15}$~km~s$^{-1}$ (68\%)
for a `shallow' $\alpha^{(e)}=-0.54$. Finally, from the relative
frequencies of doubly-imaged sources and quadruply-imaged sources, 
we find that a mean projected mass ellipticity of early-type galaxies 
$\bar{\epsilon}_{\rm mass} = 0.42$ with a 68\% lower limit of 0.28 assuming
equal frequencies of oblates and prolates.
\end{abstract}

\begin{keywords}
gravitational lensing - cosmological parameters - galaxies: structure
- galaxies: kinematics and dynamics - galaxies: halos - methods: statistical
\end{keywords}

\section{INTRODUCTION}
\setcounter{figure}{0}
\setcounter{table}{0}

Analysing the statistical properties of strong gravitational lensing in a 
sample of cosmologically distant sources provides a method for constraining
cosmological parameters (e.g.\ Turner 1990; Fukugita, Futamase, \& Kasai 1990;
Fukugita \& Turner 1991; Fukugita et al.\ 1992; Carroll, Press, \& Turner 
1992) as well as for constraining global properties of galaxy populations
including evolutions (e.g.\ Maoz \& Rix 1993; Kochanek 1993; Mao \& Kochanek
1994; King \& Browne 1996). The observed statistical properties of 
gravitational lensing in a sample are the total rate of multiple-imaging, 
the image separations, the lens redshifts, the source redshifts, and the 
image multiplicities. These properties depend not only on cosmological 
parameters but also on the following properties of galaxies and sources 
(e.g.\ Turner, Ostriker, \& Gott 1984; Turner 1990; Fukugita \& Turner 1991; 
Mao 1991; Kochanek 1993; Maoz \& Rix 1993; Wallington \& Narayan 1993; 
Mao \& Kochanek 1994; King \& Browne 1996; Kochanek 1996a,b): (1) the galaxy 
luminosity functions per morphological type (and their evolutions of any 
relevance to lensing) and the projected mass distribution in the inner 
cylindrical region of a galaxy of given luminosity and morphological type; 
and (2) the distribution of the sources in redshift and flux in the 
observational selection waveband (e.g.\ radio). 
Because of the dependences of the statistical properties of gravitational
lensing on the above multiple factors, analyses of statistical lensing can 
be used for various purposes. On the other hand, in order to derive reliable 
results from an analysis of statistical lensing, it is vital to properly take
into account all the factors through a statistical lensing model based on 
reliable data.

In this paper we perform extensive analyses of statistical lensing based on
more reliable sets of data than were available previously and using a more 
realistic statistical lensing model. Essential results on cosmological 
parameters from (part of) the analyses are reported in Chae et al.\ (2002).
However, in this paper we present not only the details of the analyses but
also comprehensive results of the analyses for a broader range of parameters.
In particular, we derive and discuss the global properties of the galaxies.
In comparison to previous works on statistical lensing
that attempted to put limits on cosmological parameters
(e.g.\ Kochanek 1996a; Falco, Kochanek, \& Mu\~{n}oz 1998;
Chiba \& Yoshii 1999; Helbig et al.\ 1999; Cooray 1999), 
the following aspects of this work can be pointed out:
\begin{enumerate}
\item This work is based on the final Cosmic Lens All-Sky Survey 
(CLASS: Myers et al.\ 2003; Browne et al.\ 2003) 
statistical sample of 8958 radio sources [which includes 
Jodrell-Bank--VLA Astrometric Survey (JVAS) sources as a subsample] 
containing 13 multiply-imaged sources, which is the largest statistical 
sample that satisfies well-defined observational selection criteria.

\item This work is based on estimates of luminosity functions of galaxies per
morphological type that have been derived taking into account latest results 
from large-scale observations of galaxies particularly including the Two 
Degree Field Galaxy Redshift Survey (2dFGRS) and the Sloan Digital Sky Survey
(SDSS).

\item We adopt a statistical lensing model in which we not only use
elliptical projected densities for galaxies but also incorporate intrinsic 
shapes of galaxies. The adopted galactic 
model is the axisymmetric (i.e.\ oblate and prolate)
singular isothermal ellipsoid (SIE) with a distribution
function of the form ${\cal F} = {\cal F}(E,L_z)$, where $E$ and $L_z$ are 
respectively the relative energy and the angular momentum component parallel
to the symmetry axis. 

\item This work is intended to provide robust constraints in cosmological
parameter space. We make the least possible prior assumptions on the 
parameters of the assumed statistical lensing model and constrain the 
parameters self-consistently from the data of statistical lensing. 

\item This work is intended to be an investigation of global properties of 
galaxy populations; namely, characteristic velocity dispersion(s) for the 
early-type (and the late-type) galaxy population(s), mean apparent mass axial 
ratio of galaxies, and intrinsic shapes of galaxies. These investigations
are made possible by the use of the (3-dimensional) ellipsoidal lens model
[the above point~(iii)].
\end{enumerate}

This paper is organized as follows. In section~2, we build up the statistical
lensing model based on the singular isothermal ellipsoid (SIE) mass model
for a galaxy, including the calculation of the cross sections for 
double-imaging and quadruple-imaging as functions of line-of-sight velocity
dispersion, apparent ellipticity and assumed intrinsic shape distributions, 
and the calculation of magnification biases for double-imaging and 
quadruple-imaging. In section 2, we also define the likelihood function 
for the data given the model of statistical lensing. 
In section~3, we summarize the input data for the analyses including the 
final CLASS statistical sample, distribution of the CLASS sources
in redshift and radio flux density, and galaxy luminosity functions 
per morphological type. The results of fitting the model to the data
are presented in section~4. In section~5, we discuss the derived results 
including comparison with previous results from statistical lensing and 
examination of possible systematics that could affect the derived results.
In section~6, we discuss future prospects of statistical lensing.
Conclusions are given in section~7.

\section{MODEL AND METHOD}
\subsection{Lens model: the singular isothermal ellipsoid}
The isothermal mass profile (i.e.\ the density $\rho \sim r^{-2}$) has been 
widely used for analyses of statistical lensing (e.g.\ Turner, Ostriker, \& 
Gott 1984; Fukugita \& Turner 1991; Kochanek 1993; Maoz \& Rix 1993;
Wallington \& Narayan 1993; Mao \& Kochanek 1994; King \& Browne 1996; 
Kochanek 1996a,b; Falco, Kochanek, \& Mu\~{n}oz 1998; Chiba \& Yoshii 1999; 
Helbig et al.\ 1999; Cooray 1999; Rusin \& Tegmark 2001), 
because of the analytical tractability of the isothermal profile and because
the projected surface density of the inner cylindrical regions of lensing 
galaxies probed by strong gravitational lensing are either consistent with
or not too different from the isothermal profile.\footnote{The reader is 
referred to numerous examples in the literature [see, e.g., Treu \& Koopmans 
(2002) and Koopmans \& Treu (2003) for very recent examples].}
Isothermal mass models would, in general, be too simplified to allow
us to do accurate modelling of individual lenses (see, e.g., Mu\~{n}oz, 
Kochanek, \& Keeton 2001; Chae 2002), but they should suffice
at least as first-order approximations to the mean properties of galaxies 
relevant to statistical lensing. Thus, following the literature on
statistical lensing, we adopt the isothermal mass profile for our analyses.
However, we use a more general isothermal model that incorporates not only a 
galaxy's apparent axial ratio but also its intrinsic shape.\footnote{All the
previous works on statistical lensing that intended to put limits on 
cosmological parameters used an isothermal sphere, while the works that
focused on studying the relative frequencies of image multiplicities used 
an elliptical lens model (e.g.\ King \& Browne 1996; Kochanek 1996b; 
Rusin \& Tegmark 2001).}

\subsubsection{The surface density, dynamical normalisations, and cross 
sections}
The projected surface density of the SIE lens $\Sigma^{\rm SIE}(x,y)$ can be 
written in units of the critical surface density $\Sigma_{\rm cr}$ by
\begin{eqnarray}
\kappa^{\rm SIE}(x,y) & \equiv &
\frac{1}{\Sigma_{\rm cr}} \Sigma^{\rm SIE}(x,y) \nonumber \\
 & \equiv & \frac{4\pi G}{c^2} R_H \frac{\hat{D}(0,z_l) 
\hat{D}(z_l,z_s)}{\hat{D}(0,z_s)} 
  \Sigma^{\rm SIE}(x,y) \nonumber \\
 & \equiv & \frac{r_{\rm cr}}{2}
  \frac{\sqrt{f} \lambda(f)}{\sqrt{x^2 + f^2 y^2}},
\end{eqnarray}
where $f$ is the minor-to-major axial ratio of the surface density and is 
equal to $1- \epsilon$ (where $\epsilon$ is the ellipticity), $R_H$ is the 
Hubble length $c/H_0$, $\hat{D}(z_1,z_2)$ is the angular diameter distance 
between redshift $z_1$ and $z_2$ ($z_2 > z_1$) in units of $R_H$, and $z_l$ 
and $z_s$ are respectively the lens and the source redshifts. In equation~(1),
$r_{\rm cr}$ is the critical radius for the singular isothermal sphere (SIS)
(i.e.\ for $f=1$) and is given by
\begin{equation}
r_{\rm cr} = 4\pi 
\left(\frac{\sigma}{c}\right)^2 R_H
 \frac{\hat{D}(0,z_l) \hat{D}(z_l,z_s)}{\hat{D}(0,z_s)},
\end{equation}
where $\sigma$ is the line-of-sight velocity dispersion of the 
galaxy.\footnote{Here we are considering a single-component mass model for
the total mass distribution of the galaxy.} For the case of the SIS, only 
double imaging is allowed and  the cross section for the double imaging is 
$\pi r_{\rm cr}^2$. For $f \neq 1$, 
quadruple imaging (and triple imaging for a sufficiently
flattened surface density; see below) is possible in addition to double
imaging, and the relations between the multiple-imaging cross sections
and the line-of-sight velocity dispersion of the galaxy are not unique but 
depend on the intrinsic shape, the viewing direction, and 
the distribution function of the galaxy. This dependence is 
parametrised by $\lambda(f)$ (`dynamical normalisation') in
equation~(1). In our parametrisation of the SIE in equation~(1),
$\lambda(f=1) =1$ and the case $\lambda(f) = 1$ for any $f$ corresponds to the
normalisation in which the mass within the area enclosed by an equi-density 
ellipse is independent of $f$ (see Kormann, Schneider, \& Bartelmann 1994). 
As shown in Kormann, Schneider, \& Bartelmann (1994), for 
$f < f_0 (\approx 0.3942)$ triple imaging with comparable brightnesses
for all three images is possible due to a naked cusp radial caustic.
The cross section for the naked-cusp triple-imaging is expected to be small
and no case has been found from the CLASS (Browne et al.\ 2003). Hence we
ignore the possibility in the following analyses.

Let $\hat{s}_2 (f)$ and $\hat{s}_4 (f)$ denote respectively the double imaging
and the quadruple imaging cross sections in units of $r_{\rm cr}^2$ for the
case of $\lambda(f)=1$ for any $f$. Then, for any given
$\lambda(f)$, the double  (quadruple) imaging cross section $s_2(f)$ 
[$s_4(f)$] in units of $r_{\rm cr}^2$ is given by 
$s_2(f)=[\lambda(f)]^2 \hat{s}_2 (f)$ \{$s_4(f)=[\lambda(f)]^2\hat{s}_4(f)$\}.
The cross sections for the double imaging and the quadruple imaging for 
$\lambda(f)=1$ are respectively given by  (Kormann et al.\ 1994)
\begin{eqnarray}
\hat{s}_4 (f) & = &  \frac{4 f}{1-f^2}  \nonumber \\
 &  & \times \int_f^{1} \left(\frac{\sqrt{1-x^2}}{x} - \arccos x \right) 
  \frac{\sqrt{x^2-f^2}}{x^2} dx,
\end{eqnarray}
and
\begin{equation}
\hat{s}_2(f) = \pi-2\hat{s}_4(f)
\end{equation}
which are valid for $f \ge f_0$. Notice that the cross sections given by 
equations~(3) and (4) are for all possible image magnification ratios of the 
multiple images. However, for the CLASS statistical sample (section~3.1) the 
observational selection limit on the image magnification ratio for the
double imaging is ${\mathcal R} > 0.1$ where ${\mathcal R}$ is the ratio
of the flux densities of the fainter to the brighter component.
Thus, the double imaging cross section that satisfies the observational limit
on the image magnification ratio is less than that given by equation~(4) but
it can be calculated numerically through a Monte Carlo method. 
The total cross section for multiple imaging for the SIE 
is given, in units of $r_{\mbox{\scriptsize cr}}^2$, by
\begin{eqnarray}
s_{\mbox{\scriptsize tot}}(f) & = & s_2(f) + s_4(f)  \nonumber \\
 & = & [\lambda(f)]^2 \hat{s}_{\mbox{\scriptsize tot}}(f) \nonumber \\
 & = & [\lambda(f)]^2 [\hat{s}_2 (f) + \hat{s}_4 (f)].
\end{eqnarray}

The rest of this subsection is devoted to
the calculation of the dynamical normalisation factor 
$\lambda(f)$ in equation~(1). For the calculation of the dynamical 
normalisation, we need to calculate the average of the line-of-sight velocity
dispersions for an ensemble of models that are allowed by the axial ratio $f$
of the surface mass density (equation~1). In general, calculations of  
velocity dispersions of a galaxy require numerically constructing the orbits 
of the particles in the galaxy that satisfy physical distribution functions 
${\cal F} \ge 0$.\footnote{In this paper, we use the notation ${\cal F}$
for the distribution function, since the notation $f$ is used to denote the 
surface density axial ratio.}
However, in this paper we consider the cases where the velocity 
dispersions can be calculated analytically. These are the axisymmetric
models, namely the oblate and the prolate spheroid, with distribution
functions of the form ${\cal F} = {\cal F}(E,L_z)$, where $E$ and $L_z$ 
are respectively the relative energy and the angular momentum component 
parallel to the symmetry axis. The axisymmetric singular isothermal mass 
model can be written in the cylindrical coordinates $(R,Z)$ and in the polar
coordinates $(r,\vartheta)$ by
\begin{eqnarray}
\rho & = & \rho_0 \frac{1}{R^2+q^{-2}Z^2} = 
 \frac{\sigma_0^2}{2\pi G} \frac{1}{R^2+q^{-2}Z^2} \nonumber \\
 & = & \frac{\sigma_0^2}{2\pi G} 
\frac{1}{r^2(\sin^2\vartheta+q^{-2}\cos^2\vartheta)},
\end{eqnarray}
where $q$ is the intrinsic axial ratio of the spheroid. The oblate and the 
prolate spheroids respectively correspond to $q < 1$ and $q > 1$. For $q=1$ 
(i.e.\ the SIS case), the parameter $\sigma_0$ is identical to the 
one-dimensional velocity dispersion $\sigma$ (equation~2).
Under the assumption that the distribution function has the form 
${\cal F} = {\cal F}(E,L_z)$, the $Z$-component $\sigma_Z^2$ and the 
$R$-component $\sigma_R^2$ of the second velocity moments 
are identical [see, e.g., Binney \& Tremaine 1987, equation (4-175)]. 
The azimuthal component of the velocity distribution at a point
includes streaming rotation motion as well as random motions. We use the mean
squared azimuthal velocity, to be denoted by $\sigma_\varphi'^{2}$, for our
dynamical normalisation calculations since measured line-of-sight velocity
dispersions over a large aperture include contributions from any streaming 
motions. For the axisymmetric scale-free isothermal model given by 
equation~(6), the second velocity moments can be calculated analytically 
and the expressions include only elementary functions 
(see, e.g., van der Marel 1994; Qian et al.\ 1995).
The second velocity moments are given for the oblate and
the prolate cases as follows: (1) For $q < 1$, we have
\begin{eqnarray}
\sigma_R^2(\vartheta) & = & \sigma_Z^2(\vartheta) \nonumber \\
  & = & \sigma_0^2 (1+q^{-2}\cot^2\vartheta)
 \left(\frac{q}{q'}\right)^2 \nonumber \\
 & & \times \left[\arctan^2\left(\frac{q'}{q}\right)
  - \arctan^2\left(\frac{q'}{q}\cos\vartheta\right) \right],
\end{eqnarray}
\begin{equation}
\sigma_\phi'^{2} (\vartheta) = 
    2 \sigma_0^2 \frac{q}{q'} \arcsin q'-\sigma_R^2(\vartheta);
\end{equation}
and (2) for $q > 1$, we have
\begin{eqnarray}
\sigma_R^2(\vartheta) & = & \sigma_Z^2(\vartheta) \nonumber \\
 & = & \frac{\sigma_0^2}{4} 
(1+q^{-2}\cot^2\vartheta) \left(\frac{q}{q'}\right)^2 \nonumber \\
 &  & \times \left[ \ln^2\left(\frac{q+q'}{q-q'}\right) 
 - \ln^2\left(\frac{q+q'\cos\vartheta}{q-q'\cos\vartheta}\right) \right],
\end{eqnarray}
\begin{equation}
\sigma_\varphi'^{2} (\vartheta) = \sigma_0^2 \frac{q}{q'} \ln 
\left(\frac{q+q'}{q-q'}\right) - \sigma_R^2(\vartheta). 
\end{equation}
As above, throughout we use the notation $q' \equiv |1-q^2|^{1/2}$. 

We calculate the dynamical normalisations for the oblate and the prolate cases
using the above equations (7), (8), (9) and (10).
\newline
{\it a. The oblate case ($q < 1$)---}The radial component $\sigma_R^2$ and
the azimuthal component $\sigma_\varphi'^2$ of the second velocity moments in
the cylindrical coordinates can be used to obtain a Cartesian component 
$\sigma_X^2$ that is perpendicular to the symmetry
axis, namely we have
\begin{equation}
\sigma_X^2 = \sigma_R^2 \cos^2\varphi + \sigma_\varphi'^2 \sin^2\varphi.
\end{equation}
The mass-weighted average of a second velocity moment is defined by
\begin{equation}
\overline{\sigma^2} \equiv \frac{1}{\int\rho(r^2\sin\vartheta dr d\vartheta 
d\varphi)} 
\int \rho \sigma^2 (r^2\sin\vartheta dr d\vartheta d\varphi).
\end{equation}
Using the above definition (equation~12), we obtain for the mass-weighted
averages of the $X$ and the $Z$ components respectively in units of 
$\sigma_0^2$
\begin{equation}
W_X^{\rm obl}(q) \equiv 
\overline{\sigma_X^2}/\sigma_0^2 = \frac{q}{q'} \arcsin q',
\end{equation}
and
\begin{eqnarray}
W_Z^{\rm obl}(q) & \equiv &
\overline{\sigma_Z^2}/\sigma_0^2  \nonumber \\
 & = & \frac{1}{2} \frac{q}{q'}
\left( \arctan\frac{q'}{q} \right)^{-1} 
\int_0^{\pi}d\vartheta \frac{1}{\sin\vartheta} \nonumber \\
 &  & \times
\left( \arctan^2 \frac{q'}{q} - \arctan^2 \frac{q'\cos\vartheta}{q} \right).
\end{eqnarray}
Given the mass-weighted averages of the two Cartesian components, the 
line-of-sight component of the second velocity moments when the galaxy is 
viewed at an inclination angle of $i$, with $i = \pi/2$ defining the edge-on 
view, is given by
\begin{equation}
\sigma^2_{\rm los}(i) = 
 \sigma_0^2[W^{\rm obl}_Z (q) \cos^2 i + 
W^{\rm obl}_X (q) \sin^2 i].
\end{equation}
For a given axial ratio $f$ of the projected surface density given by 
equation~(1), there are infinitely many combinations of inclination angle
$i$ and intrinsic axial ratio $q$ that are consistent with the projected
axial ratio $f$. Averaging equation~(15) over this ensemble of models,
we have
\begin{equation}
\langle \sigma^2_{\rm los} \rangle = (\pi/2 -  i_{\rm min})^{-1}
\int_{i_{\mbox{\tiny min}}}^{\pi/2} \sigma^2_{\rm los}(i) di, 
\end{equation}
where $i_{\rm min}$ is the minimum possible inclination angle
for the given projected axial ratio $f$.
From the relation
\begin{equation}
f^2 = \cos^2 i + q^2 \sin^2 i = 1 - q'^2 \sin^2 i
\end{equation}
(for $q<1$), we have
\begin{equation}
i_{\rm min} = \arcsin f',
\end{equation}
where and throughout we use the notation $f' \equiv (1 - f^2)^{1/2}$.
Introducing the variable $t = \sin i$, we can write the inclination-averaged
line-of-sight velocity second moment as
\begin{eqnarray}
\langle \sigma^2_{\mbox{\scriptsize los}} \rangle (f') & = & \sigma_0^2 
(\pi/2 - \arcsin f')^{-1} \nonumber \\
 &  & \times \int_{f'}^1  \{ [W^{\rm obl}_X(q(t))-W^{\rm obl}_Z(q(t))] t^2 
   \nonumber \\ 
 &  & \hspace{4em} +W^{\rm obl}_Z(q(t)) \}\frac{dt}{\sqrt{1 -t^2}}.
\end{eqnarray}
The projected surface density for the oblate case of the axisymmetric mass 
distribution given by equation~(6) can be written as 
\begin{equation}
\Sigma^{\mbox{\scriptsize obl}} = 
\frac{\sigma_0^2}{2 G}  q  
\frac{1}{\sqrt{x^2+f^2 y^2}}
\Rightarrow
\Sigma^{\mbox{\scriptsize obl}} = 
\frac{\sigma_0^2}{2 G} \langle q \rangle 
\frac{1}{\sqrt{x^2+f^2 y^2}},
\end{equation}
where $\langle q \rangle$ is the inclination averaged value of the axial 
ratio, which can be shown to be
\begin{equation}
\langle q \rangle = \langle q \rangle (f) = 
(\pi/2 - \arcsin f')^{-1} \frac{\pi}{4} f^2 
 F\left(1,\frac{1}{2};2;f^2\right),
\end{equation}
where $F$ is a hypergeometric function. Finally,
comparing equation~(20) with equation~(1) in conjunction with the 
relations given by equations (19) and (21), we obtain for the dynamical
normalisation of the singular isothermal oblate spheroid with a distribution
function of the form ${\mathcal F} = {\mathcal F}(E,L_z)$
\begin{eqnarray}
\lambda^{\mbox{\scriptsize obl}}(f) & = &
\frac{\pi}{4}f^{3/2} F\left(1,\frac{1}{2};2;f^2\right) \nonumber \\
 &  & \times \left[ \int_{f'}^1  
\{ [W^{\rm obl}_X(q(t))-W^{\rm obl}_Z(q(t))] t^2  \right. 
   \nonumber \\  &  & \left.   \hspace{4em}
 +W^{\rm obl}_Z(q(t))\}\frac{dt}{\sqrt{1 -t^2}}
 \right]^{-1}.
\end{eqnarray}
{\it b. The prolate case ($q > 1$)---} A procedure similar to the above
procedure for the oblate case can be followed using equations (9) and (10)
to obtain the dynamical normalisation for the prolate spheroid. We point out
the differences in the procedures between the oblate and the prolate cases and
then list the calculated results. First, the projected axial ratio $f (< 1)$
is related to the intrinsic axial ratio $q (> 1)$ via
\begin{equation}
f^2 = (\cos^2 i + q^2 \sin^2 i)^{-1} = (1 + q'^2 \sin^2 i)^{-1}.
\end{equation}
 Second, the projected surface density for the prolate case
of the axisymmetric mass distribution given by equation~(6) can be written
using equation~(23) as 
\begin{eqnarray}
\Sigma^{\mbox{\scriptsize prol}}  = 
 \frac{\sigma_0^2}{2 G} f  q  
 \frac{1}{\sqrt{x^2+f^2 y^2}} \Rightarrow &  &  \nonumber \\
\Sigma^{\mbox{\scriptsize prol}}   = 
 \frac{\sigma_0^2}{2 G} f \langle q \rangle 
 \frac{1}{\sqrt{x^2+f^2 y^2}}.  &   &
\end{eqnarray}
Finally, the requirement of the positivity of the second velocity moments
 shows that the intrinsic axial ratio $q$ cannot be 
greater than $q_{\mbox{\scriptsize max}} = 3.46717\ldots$ 
and thus the minimum inclination angle $i_{\mbox{\scriptsize min}}$ 
for the given projected axial ratio $f$ is given by
\begin{equation}
i_{\mbox{\scriptsize min}}=\arcsin t_{\mbox{\scriptsize min}} =\arcsin \left(
\frac{1}{\sqrt{q_{\mbox{\scriptsize max}}^2 - 1}} \frac{f'}{f} \right).
\end{equation}
Unfortunately, even for $q < q_{\mbox{\scriptsize max}}$, non-negativity of
the distribution function is not guaranteed. In other words, our adopted 
(simple-minded) dynamical model for the singular isothermal prolate spheroid
has an unphysical feature.  Nonetheless,
our adopted dynamical model was sometimes used in the literature as
a simple approximation. Furthermore, the construction of a physical dynamical 
model in conjunction with a prolate mass model that would mimic the SIE is 
beyond the scope of this paper.
Bearing the above in mind, we now give the calculated results.
The dynamical normalisation of the singular isothermal prolate spheroid with
a distribution function of the form ${\mathcal F} = {\mathcal F}(E,L_z)$ 
is given by
\begin{eqnarray}
\lambda^{\mbox{\scriptsize prol}}(f) &  = &
 \left( \int_{t_{\rm min}}^1 
 \sqrt{\frac{t^2f^2 + f'^2}{f (1-t^2)}} \frac{dt}{t} \right) \nonumber \\ 
 &  & \times \left[ \int_{t_{\rm min}}^1 
\{ [W^{\rm prol}_X(q(t))-W^{\rm prol}_Z(q(t))] t^2  \right. 
   \nonumber \\  &  & \left.   \hspace{5em}
 +W^{\rm prol}_Z(q(t))\}\frac{dt}{\sqrt{1 -t^2}} \right]^{-1}. 
\end{eqnarray}
where
\begin{equation}
W^{\mbox{\scriptsize prol}}_X(q) = \frac{1}{2}\frac{q}{q'} 
\ln \frac{q+q'}{q-q'},
\end{equation}
and
\begin{eqnarray}
W^{\mbox{\scriptsize prol}}_Z(q) & = & \frac{1}{4}\frac{q}{q'}
\left( \ln \frac{q+q'}{q-q'} \right)^{-1} 
\int_0^{\pi} d\vartheta \frac{1}{\sin\vartheta} \nonumber \\
 &  & \times \left[ \left(\ln \frac{q+q'}{q-q'} \right)^2 - 
 \left(\ln \frac{q+q'\cos\vartheta}{q-q'\cos\vartheta} \right)^2 \right].
\end{eqnarray}
\begin{figure}
\begin{center}
\setlength{\unitlength}{1cm}
\begin{picture}(8,8)(0,0)
\put(-1.,-2.5){\includegraphics{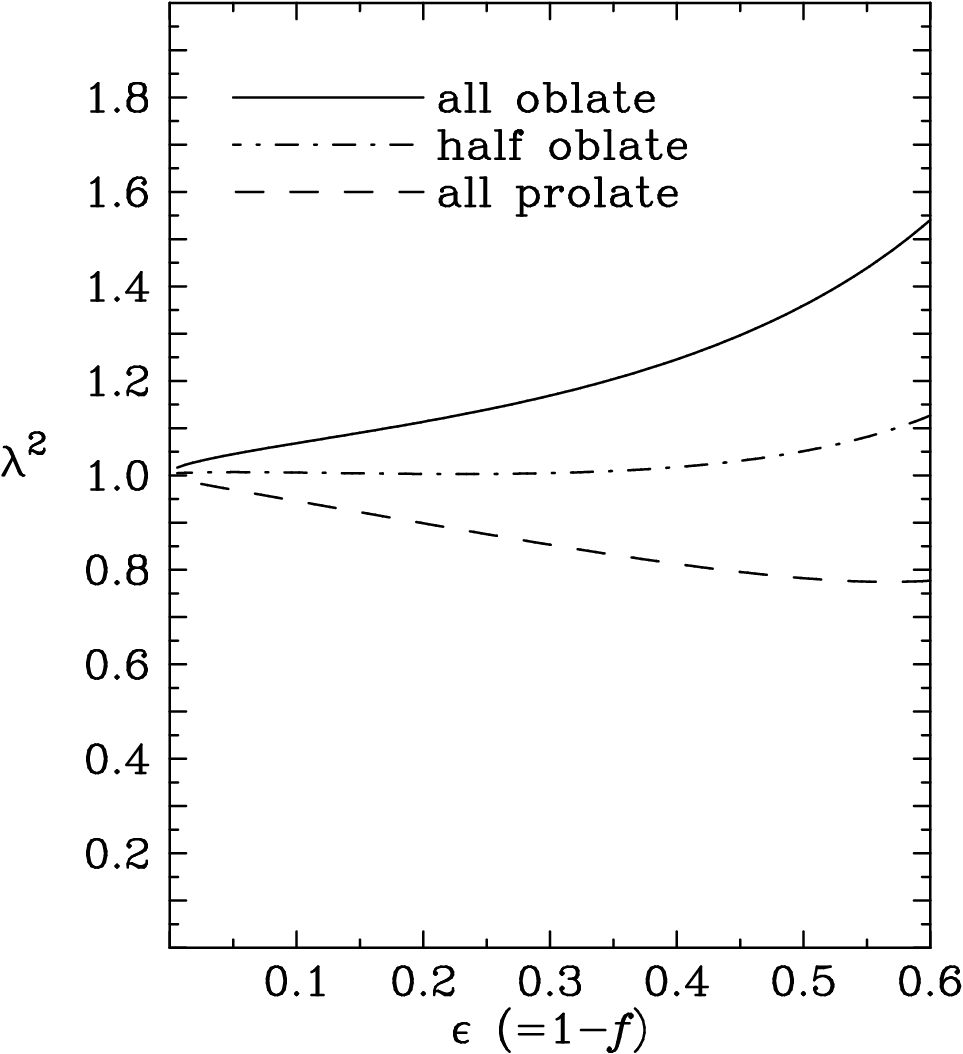}}
\end{picture}
\caption{
Functional behaviour of the lens `dynamical normalisation' squared. 
We consider three cases of the dynamical normalization, namely 
$\lambda(f) = \lambda^{\mbox{\scriptsize obl}}(f)$,
 $\lambda(f) = 0.5\lambda^{\mbox{\scriptsize obl}}(f)+
       0.5\lambda^{\mbox{\scriptsize prol}}(f)$,
and $\lambda(f) = \lambda^{\mbox{\scriptsize prol}}(f)$, where 
$\lambda^{\mbox{\scriptsize obl}}(f)$ and 
$\lambda^{\mbox{\scriptsize prol}}(f)$ are respectively the dynamical 
normalisations for the oblate and the prolate scale-free isothermal spheroids.
}
\label{}
\end{center}
\end{figure}
{Fig.}~1 shows the functional behaviour of the dynamical normalisation 
squared for the following three cases; 
 $\lambda(f) = \lambda^{\mbox{\scriptsize obl}}(f)$,
 $\lambda(f) = 0.5\lambda^{\mbox{\scriptsize obl}}(f)+
       0.5\lambda^{\mbox{\scriptsize prol}}(f)$,
and $\lambda(f) = \lambda^{\mbox{\scriptsize prol}}(f)$, where 
$\lambda^{\mbox{\scriptsize obl}}(f)$ and 
$\lambda^{\mbox{\scriptsize prol}}(f)$ are respectively given by 
equations~(22) and (26) (see section~2.1.1).

\subsubsection{Galaxy luminosity functions, magnification biases, and
lensing probabilities}
We assume that galaxies of a given morphological type are 
uniformly distributed in comoving space and distributed in luminosity
following a Schechter (1976) luminosity function and that the luminosity of 
a galaxy has a power-law relation to its line-of-sight velocity dispersion
(i.e.\ the Faber-Jackson relation for early-type galaxies and the
Tully-Fisher relation for late-type galaxies).
The differential probability for a source with redshift $z_s$ and 
flux density $S_{\nu}$ to be multiply-imaged with image multiplicity $m$
and image separation $\Delta\theta$ to $\Delta\theta+d(\Delta\theta)$,
due to a distribution of intervening galaxies of morphological type 
$g$ (where $g= e$ for early-type galaxies and $s$ for late-type galaxies) 
at redshifts $z$ to $z+dz$ modelled by SIEs and described by a Schechter
luminosity function, is given by
\begin{eqnarray}
\frac{d^2p_m^{(g)}(z,\Delta\theta;z_s,S_{\nu})}{dz d(\Delta\theta)} 
& = & s_m(f) 8 \pi^2 \gamma  R_H^3  \nonumber \\
&  &\times n_{*}(z)(1+z)^3\left[\frac{\sigma_*(z)}{c} \right]^4 \nonumber \\
&  & \times \left|\frac{d\ell}{dz}\right|
\left[\frac{\hat{D}(0,z) \hat{D}(z,z_s)}{\hat{D}(0,z_s)}\right]^2 \nonumber \\
&  & \times \frac{1}{\Delta\theta_*(z)} 
\left[ \frac{\Delta\theta}{\Delta\theta_*(z)} 
\right]^{(\alpha\gamma+\gamma+2)/2} \nonumber \\
  &  & \times  
\exp \left[ - \left(\frac{\Delta\theta}{\Delta\theta_*(z)} 
\right)^{\gamma/2} \right]  \nonumber \\
  &  & \times   B_m(z_s, S_{\nu}), 
\end{eqnarray}
where $s_m(f)$ (hereafter, $m= 2, 4$) is the cross section in units of
$r_{\rm cr}^2$ (equation~2) for multiple imaging with image multiplicity $m$ 
(section~2.1.1), $\Delta\theta_*(z)$ is the characteristic image separation
(see below), and $B_m(z_s, S_{\nu})$ is the magnification bias (see below).
In equation~(29), $|d\ell/dz|$, where $\ell=H_0\times$(proper time), is 
given by
\begin{eqnarray}
\left|\frac{d\ell}{dz}\right| & = & (1+z)^{-1} 
 [ \Omega_{\mbox{\scriptsize m}} (1+z)^3 \nonumber \\
 &  & \hspace{2em} + (1-\Omega_{\mbox{\scriptsize m}} - 
   \Omega_{\Lambda}) (1+z)^2 + \Omega_{\Lambda}]^{-1/2}
\end{eqnarray}
for a cosmology with a classical cosmological constant.
In equation~(29), we have also used the following notations:
the Schechter (1976) function 
\begin{equation}
\frac{dn(L,z)}{dL} = \frac{n_*(z)}{L_*(z)} 
\left[\frac{L}{L_*(z)}\right]^{\alpha}  \exp[-L/L_*(z)],
\end{equation} 
where $\alpha$ is the faint-end slope, $L_*(z)$ is the characteristic
luminosity, and
$n_*(z)$ is the characteristic comoving number density which can be written as
\begin{equation}
n_* (z) = n_{*,0} e_n (z).
\end{equation}
The luminosity of a galaxy is assumed to be correlated with its line-of-sight
velocity dispersion by 
\begin{equation}
\frac{L}{L_*(z)} = 10^{0.4[M_*(z)-M]} = 
\left[\frac{\sigma}{\sigma_*(z)}\right]^{\gamma}
\end{equation}
where $M_*(z)$ and $\sigma_*(z)$ are respectively the characteristic absolute
magnitude and the characteristic velocity dispersion corresponding to
the characteristic luminosity $L_*(z)$, and $\gamma$ is the
Faber-Jackson or Tully-Fisher exponent. We write the characteristic
velocity dispersion as 
\begin{equation}
\sigma_* (z) = \sigma_{*,0} e_v(z).
\end{equation}
In equation~(29), the characteristic image separation $\Delta\theta_*(z)$
by an $L_*$ galaxy is given by
\begin{equation}
\Delta\theta_*(z) = \lambda(f) 8 \pi \frac{\hat{D}(z,z_s)}{\hat{D}(0,z_s)} 
\left[ \frac{\sigma_* (z)}{c} \right]^2,
\end{equation}
where $\lambda(f)$ is the dynamical normalisation.
Notice that the characteristic image separation $\Delta\theta_*(z)$ 
(equation~35) scales linearly with the dynamical normalisation $\lambda(f)$,
while the multiple imaging cross sections $s_m(f)$ scale quadratically with 
$\lambda(f)$ (section~2.1.1).
In equations (32) and (34), parameters $e_n(z)$ and $e_v(z)$ respectively 
represent the evolutions of the number density and the velocity dispersion of
galaxies. 

The magnification bias $B_m(z_s,S_{\nu})$ in equation (29), namely the factor
by which the multiply-imaged sources are overrepresented compared with the
unlensed sources in a flux limited sample because the multiply-imaged sources
come from intrinsically fainter populations compared with the unlensed ones 
of the same apparent brightness (see, e.g., Maoz \& Rix 1993; 
King \& Browne 1996), is given by
\begin{eqnarray}
B_m(z_s,S_{\nu}) & = &
\left[ \int_{\mu_{m,\mbox{\scriptsize min}}}^{\mu_{m,\mbox{\scriptsize max}}} 
\left|\frac{dN_{z_s}(>S_{\nu}/\mu_m)}{dS_{\nu}}
 \frac{dP_m(>\mu_m)}{d\mu_m} \right| \right. \nonumber \\
 &  & \hspace{1em} \left. \times  \frac{1}{\mu_m} d\mu_m \right] 
\left| \frac{dN_{z_s}(>S_{\nu})}{dS_{\nu}} \right|^{-1}
\end{eqnarray}
($m=$ 2, 4).
Here $N_{z_s}(>S_{\nu})$ is the intrinsic number-flux density relation, i.e.\
the integrated source counts as a function of flux density $S_{\nu}$,
for the source population at redshift $z_s$. Parameter $\mu_m$ 
denotes the total magnification for multiple imaging with image 
multiplicity $m$, namely the sum of the absolute magnifications of the $m$ 
images. $P_m(>\mu_m)$ is the fraction of the multiple imaging cross section
with image multiplicity $m$ that have total magnifications greater than
$\mu_m$, for a given lens model. Hence $|dP_m(>\mu_m)/d\mu_m|$ is the 
magnification probability distribution above a minimum possible
magnification $\mu_{m,{\rm min}}$ and it follows that
$P_m(>\mu_{m,{\rm min}})=1$. For example, for the SIS
the magnification probability distribution for double imaging with any
magnification ratio of the two images is given
by $|dP_2^{\rm SIS}(>\mu_2)/d\mu_2| = 8/\mu_2^3$ with the minimum
total magnification of $\mu_{2,{\rm min}} = 2$. 
The minimum and maximum total magnifications $\mu_{m,{\rm min}}$ and 
$\mu_{m,{\rm max}}$ in equation~(36) depend on observation characteristics 
as well as the lens model. For multiple imaging of a point source, 
$\mu_{\mbox{\scriptsize max}} \rightarrow \infty$. 
For the CLASS statistical sample, the observational lower limit on
the ratio of the flux densities of the fainter to the brighter image for
the doubly-imaged systems is ${\mathcal R}_{\rm min} = 0.1$. Given such
an observational limit ${\mathcal R}_{\rm min}$, the minimum total 
magnification for double imaging for the SIS model is given by
\begin{equation}
\mu_{2,{\rm min}}^{\rm SIS} = 
2 \frac{1 + {\mathcal R}_{\rm min}}{1 - {\mathcal R}_{\rm min}}.
\end{equation}
For the SIE (i.e.\ $f \neq 1$ in equation~1), we numerically calculate 
magnification probability distributions and minimum total magnifications. 
We do this by solving the simplified lens equation for the SIE 
(Kormann et al.\ 1994) for millions of
source positions randomly distributed on a source region that includes
the region enclosed by the caustics for each value of the axis ratio $f$.
{Fig.}~2(a) shows several examples of the magnification probability
distribution  $|dP_m(>\mu_m)/d\mu_m|$ for double and quadruple imaging.
{Fig.}~2(b) shows the theoretical minimum total magnifications for double and
quadruple imaging, i.e.\ for any magnification ratios, 
and the minimum total magnifications for double imaging with
the CLASS observational limit ${\mathcal R}> 0.1$.
\begin{figure*}
\begin{center}
\setlength{\unitlength}{1cm}
\begin{picture}(14,13)(0,0)
\put(-2.,13.7){\includegraphics{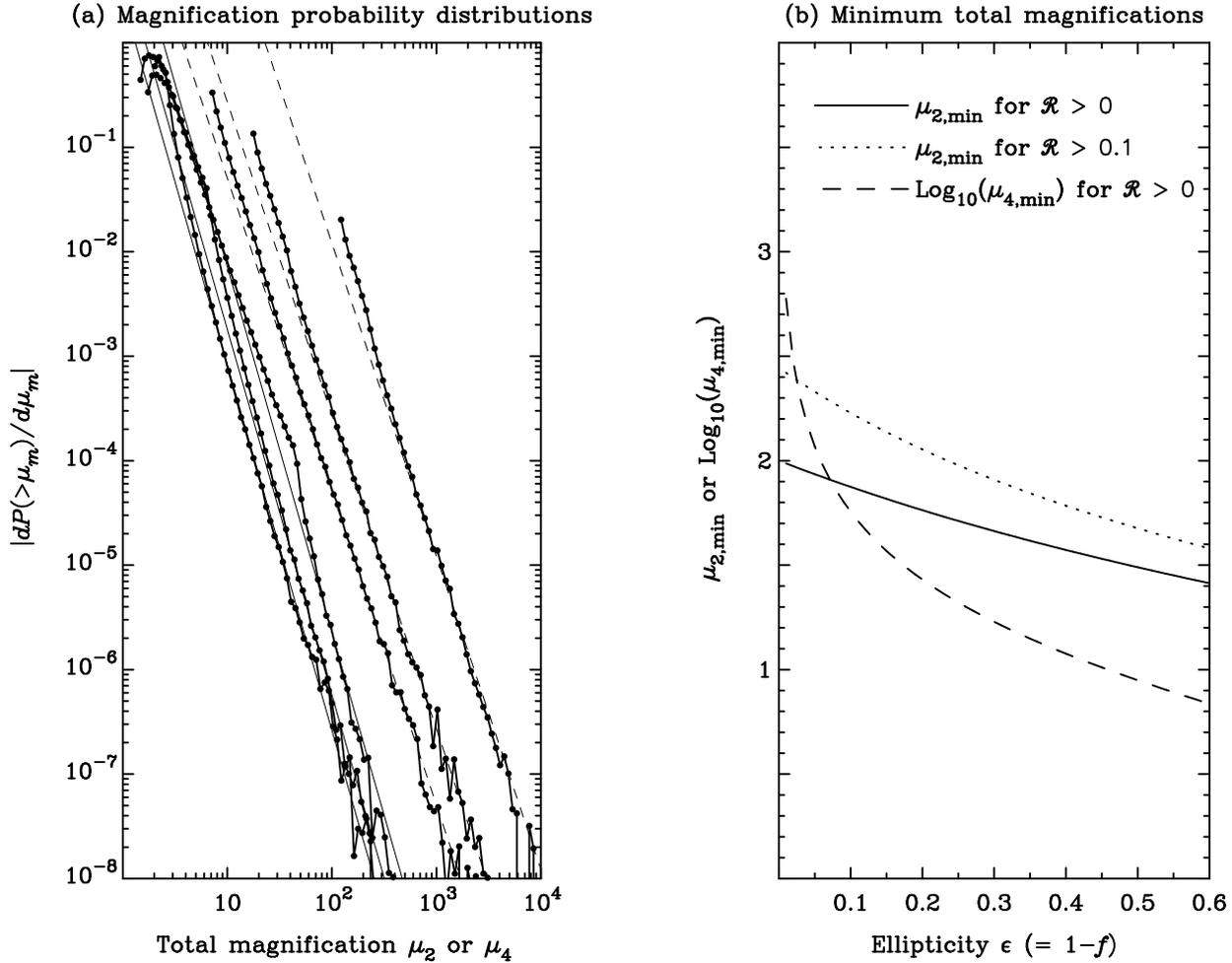}}
\end{picture}
\caption{
(a) Magnification probability distributions (MPDs). The six curves from 
right to left respectively correspond to MPDs for quadruple imaging by 
$\epsilon = 0.05$, 0.3 and 0.6 SIEs and for double imaging by 
$\epsilon = 0.05$, 0.3 and 0.6 SIEs. The thin solid lines representing 
$\mu_2^{-7/2}$ behaviours and the thin dashed lines representing $\mu_4^{-3}$ 
behaviours fit respectively well the numerical data points for double imaging
and quadruple imaging at high magnifications 
in agreement with the expectations from analytical calculations.
(b) Minimum total magnifications for double and quadruple imaging by
SIEs as a function of ellipticity $\epsilon$. The solid line shows
the minimum total magnifications for double imaging for any magnification
ratios between the two images. The dotted line shows the minimum total
magnifications for double imaging only for the fainter image to the brighter 
image magnification ratio ${\mathcal R} > 0.1$, which is the observational
limit for the final CLASS statistical sample. The dashed line shows
the minimum total magnifications for quadruple imaging for any magnification
ratios between the four images.
}
\label{}
\end{center}
\end{figure*}

The differential probability for a source to be multiply-imaged with
image multiplicity $m$ and image separation $\Delta\theta$ to 
$\Delta\theta+d(\Delta\theta)$ due to all the galaxies of type $g$ 
from $z=0$ to $z_s$, can be obtained from equation~(29):
\begin{equation}
\frac{dp_m^{(g)}(\Delta\theta;z_s,S_{\nu})}{d(\Delta\theta)} = \int_{0}^{z_s} 
\frac{d^2 p_m^{(g)}(z,\Delta\theta;z_s,S_{\nu})}{dz d(\Delta\theta)} dz.
\end{equation}
Likewise, the differential probability for a source to be multiply-imaged 
 with image multiplicity $m$ and with any image separation 
$\ge \Delta\theta_{\mbox{\scriptsize min}}$\footnote{An upper limit on
$\Delta\theta$ is not necessary since the lensing probability due to
galaxy lenses (given by equation~29) is negligibly small beyond an 
observational limit, e.g.\ several arcseconds.} due to 
galaxies of type $g$ at redshifts $z$ to $z+dz$ is given by
\begin{eqnarray}
\frac{dp_m^{(g)}(z;z_s,S_{\nu})}{dz} & = & s_m(f) 16 \pi^2 
\Gamma\left(\alpha+1+\frac{4}{\gamma}\right)
R_H^3  \nonumber \\ 
 &  & \times  n_{*} (z) (1+z)^3 \left[ \frac{\sigma_* (z)}{c} \right]^4 
 \nonumber \\ 
 &  & \times  \left|\frac{d\ell}{dz}\right|
\left[\frac{\hat{D}(0,z)\hat{D}(z,z_s)}{\hat{D}(0,z_s)}\right]^2 \nonumber \\
 &  &  \times    B_m(z_s, S_{\nu}) \nonumber \\
 &  & - \int_0^{\Delta\theta_{\mbox{\scriptsize min}}} 
\frac{d^2p_m^{(g)}(z,\Delta\theta;z_s,S_{\nu})}{dz d(\Delta\theta)} 
d(\Delta\theta),
  \nonumber \\
 &  & 
\end{eqnarray}
where the integrand in the last piece of the right-hand side is given
by equation~(29). Finally, using equation~(39) we obtain the integrated 
probability for the multiple imaging with image multiplicity $m$ due to
galaxies of type $g$: 
\begin{equation}
p_m^{(g)}(z_s,S_{\nu})=\int_{0}^{z_s}\frac{dp_m^{(g)}(z;z_s,S_{\nu})}{dz}dz.
\end{equation}

\subsubsection{Summary of model parameters}
The lensing probabilities given by equations~(29), (38), (39) and (40) are 
for a population of galaxies of a specific morphological type that are assumed
to be described by a set of Schechter parameters $n_*$, $\alpha$, and $M_*$ 
with corresponding $L_*$ and $\sigma_*$ along with the Faber-Jackson or 
Tully-Fischer exponent $\gamma$ (equation~33). We assume
two populations of galaxies separately contributing to multiple imaging, 
namely the early-type (i.e.\ ellipticals and S0s) and the late-type 
(i.e.\ spirals and other non-E/S0s) populations.
Separately taking into account the two populations is required.
This is because the early-type and the late-type galaxies are dynamically
different, satisfy respectively the Faber-Jackson and the Tully-Fisher
relations, and have different characteristics in multiple imaging:
the multiple imaging rate by the early-type population is $\approx 3$-6 times
higher and the early-type galaxies produce on average larger image splittings.

The parameters of our statistical lensing model can be summarized as follows:
(1) up to three cosmological parameters: $\bullet$ matter density 
$\Omega_{\mbox{\scriptsize m}}$, $\bullet$ density of a classical cosmological
constant $\Omega_{\Lambda}$ or dark energy density $\Omega_x$ with 
$\bullet$ its constant equation of state 
$w=p_x$(pressure)/$\rho_x$(energy~density);\footnote{Here all the 
cosmological densities refer to the fractions of the present critical 
density.} (2) six parameters for the early-type galaxy population, 
i.e.\ early-type Schechter parameters 
$n_{*,0}^{(e)}$, $\sigma_*^{(e)}$, and $\alpha^{(e)}$, 
Faber-Jackson exponent $\gamma_{\rm FJ}$, early-type mean mass ellipticity 
$\bar{\epsilon}^{(e)}$, and  relative frequency of the oblatenesses 
${\mathcal P}_{\rm obl}^{(e)}$ among early-type galaxies;
(3) six parameters for the late-type galaxy population, i.e.\ late-type 
Schechter parameters $n_{*,0}^{(s)}$, $\sigma_{*}^{(s)}$, and $\alpha^{(s)}$,
Tully-Fisher exponent $\gamma_{\rm TF}$, late-type mean mass 
ellipticity $\bar{\epsilon}^{(s)}$, and relative frequency of the 
oblatenesses ${\mathcal P}_{\mbox{\scriptsize obl}}^{(s)}$ among
late-type galaxy halos.\footnote{In this work, we are considering a 
single-component galaxy model. This means that for a spiral galaxy the mass 
model is intended to approximate the total mass distribution by the halo, the
bulge and the disk of the galaxy. Consequently, the model parameters 
(i.e., velocity dispersion and projected ellipticity) are effective quantities
corresponding to the combined lensing effects by the components of the galaxy.
Discussions on the lensing effects by the components of a spiral galaxy
can be found, e.g., in Keeton \& Kochanek (1998).}

In addition to the above parameters, our statistical lensing model includes
the parameters that can be used to model certain aspects of galaxy 
evolutions, i.e., $e_n^{(g)}(z)$ and $e_v^{(g)}(z)$ where $g=e,s$
(see~equations~32 and 34). 
In this work, we assume no evolution of the comoving number densities of
galaxies for the redshift range which is relevant to lensing by CLASS sources,
i.e., $e_n^{(g)}(z) = 1$ for $0 \le z \la 1$. Here note that any significant
evolution of the comoving number density of late-type galaxies has little
effect on the derived values of cosmological parameters.
Note also that while a significant evolution of the comoving number density
of early-type galaxies would have a significant effect on the derived values
of cosmological parameters, (our interpretation of) current observational 
evidence suggests that there has been little evolution of the comoving number
density of early-type galaxies since $z \sim 1$ (see the last paragraph of
section~3.3). However, evolution of the comoving number density of early-type 
galaxies is a main source of uncertainty and its systematic effects on
the derived values of cosmological parameters are discussed in section~5.3.3.
Since the characteristic velocity dispersions will be treated as
free parameters (section~4) and lens redshifts are intermediate, 
it should be understood that parameters $e_v^{(g)}(z)$ are irrelevant
and we are constraining parameters $\sigma_*^{(g)}$ for $0.3 \la z \la 1$. 
Finally, in addition to the cosmological parameters and the lens parameters,
there are parameters used to describe the CLASS source population, namely the
redshift distribution and the number--flux-density relation, which can be
found in section~3.2.

\subsection{Method of statistical analysis}
For a statistical sample that contains
$N_{\rm L}$ multiply-imaged sources and $N_{\rm U}$ unlensed sources, 
the likelihood $\mathcal{L}$ of the observation data
given the statistical lensing model is defined by
\begin{eqnarray}
\ln {\mathcal{L}} & = & \sum_{k=1}^{N_{\rm U}} 
\ln \left[1 - \sum_{m=2,4} p_m^{\rm(all)}(z_{s,k},S_{\nu,k}) \right] 
\nonumber \\ & & +  \sum_{l=1}^{N_{\rm L}} 
\ln \delta p_{m_l}^{\rm(one)}[(z_{l}),\Delta\theta_l;z_{s,l},S_{\nu,l}], 
\end{eqnarray}
where $\sum_{m=2,4} p_m^{\rm(all)}(z_{s,k},S_{\nu,k})$ is the sum of the 
double and quadruple imaging probabilities (equation~40) for the $k$-th 
unlensed source, and 
$\delta p_{m_l}^{\rm(one)}[(z_l),\Delta\theta_l;z_{s,l},S_{\nu,l}]$
is the differential probability (equation~29, 38, or 39) for the $l$-th
multiply-imaged source, which is a specific 
probability including the observed configuration of 
the source, i.e.\ its image
separation $\Delta\theta_l$, its image multiplicity $m_l$ 
(and its lens redshift $z_{l}$ if it is known).

In equation~(41), for the unlensed sources the probability to be used is the 
sum of the probabilities due to the early-type and the late-type populations,
namely
\begin{eqnarray}
 p_m^{\rm(all)}(z_{s,k},S_{\nu,k}) & \equiv & p_m^{(e)}(z_{s,k},S_{\nu,k}) 
 + p_m^{(s)}(z_{s,k},S_{\nu,k}),\nonumber \\
 &  & k=1,2,\ldots, N_{\rm U}.
\end{eqnarray}
For the multiply-imaged sources, the suitable differential probability can be
given by
\begin{eqnarray}
  & & \delta p_{m_l}^{\rm(one)}[(z_{l}),\Delta\theta_l;z_{s,l},S_{\nu,l}] 
 \equiv \nonumber \\ & & \hspace{4em} \omega_l^{(e)} 
\delta p_{m_l}^{(e)}[(z_{l}),\Delta\theta_l;z_{s,l},S_{\nu,l}]\nonumber \\
 & & \hspace{4em}  + \omega_l^{(s)} 
\delta p_{m_l}^{(s)}[(z_{l}),\Delta\theta_l;z_{s,l},S_{\nu,l}],\nonumber \\
 & & \hspace{4em}  l=1,2,\ldots, N_{\rm L},
\end{eqnarray}
where parameters $\omega_l^{(e)}$ and $\omega_l^{(s)}$ are the weights given
to the early-type and the late-type populations, respectively,
satisfying $\omega_l^{(e)} + \omega_l^{(s)} = 1$. If the lensing galaxy type 
is known to be an early-type (late-type), $\omega_l^{(e)} =1$ 
($\omega_l^{(s)} =1$). If the lensing galaxy type is unknown, 
we use $\omega_l^{(g)} = \delta p_{m_l}^{(g)}/
[\delta p_{m_l}^{(e)}+ \delta p_{m_l}^{(s)}]$ ($g = e$, $s$),
if the lens redshift is known, whereas we use $\omega_l^{(e)} = 0.8$ and
$\omega_l^{(s)} = 0.2$ otherwise. 

{From} equation~(41), a ``$\chi^2$'' is defined by
\begin{equation}
\chi^2 = - 2 \ln \mathcal{L} \hspace{1em} 
[+\mbox{Gaussian prior on $\sigma_*^{(s)}$}].
\end{equation}
As indicated in the bracket in equation~(44), the $\chi^2$ will 
sometimes include a Gaussian prior probability term of the late-type
characteristic velocity dispersion $\sigma_*^{(s)}$.
We determine `best-fit' model parameters by minimizing the $\chi^2$
(equation~44) and obtain confidence limits on the model parameters using 
the usual $\Delta\chi^2$ ($\equiv \chi^2 - \chi^2_{\rm min}$)-static, where 
$\chi^2_{\rm min}$ is the global minimum $\chi^2$ for the best-fit parameters.

\section{INPUT DATA}
In this section, we summarize data both from CLASS observations and recent 
large-scale observations of galaxies that will be used as the input for the 
statistical lensing model described in section~2. In section~3.1, 
we briefly review the definition
of the final CLASS statistical sample that satisfy well-defined
observational selection criteria (Browne et al.\ 2003) and summarize
the properties of the multiply-imaged sources in the sample.
In section~3.2, we summarize observational information on flat-spectrum 
radio sources including that from CLASS observations and estimate the 
distribution of CLASS radio sources in redshift and flux density.
In section~3.3, we survey recent observational results for galaxy
luminosity functions and obtain estimates of galaxy luminosity 
functions per morphological type based on the data from the literature.

\subsection{The final CLASS statistical sample}
The Cosmic Lens All-Sky Survey is the largest completed (radio-selected) 
galactic mass scale gravitational lens search project.
The CLASS project along with its predecessor project the Jodrell-Bank--VLA
Astrometric Survey (JVAS; see, e.g., King et al.\ 1999) identified 22 
multiply-imaged systems out of a total of 16521 radio sources targeted 
(Myers et al.\ 2003; Browne et al.\ 2003). Out of the entire CLASS sample
including the JVAS sources, a subsample of 8958 sources containing 13 
multiply-imaged systems constitutes a statistical sample that satisfies
well-defined observational selection criteria (Browne et al.\ 2003). 
The final CLASS statistical sample is the largest sample that
can be used for the purpose of statistical analyses of gravitational lensing.
The reader is referred to Myers et al.\ (2003) and Browne et al.\ (2003)
for the description of the CLASS observations including the selection
of the targets, the gravitational lens candidate selection process, 
the follow-up observations of the lens candidate sources, and the properties
of the finally confirmed individual multiply-imaged systems. 
{From} Browne et al.\ (2003), the final CLASS statistical sample 
satisfies the following well-defined selection criteria that
are observationally reliable:
\begin{enumerate}
\item The spectral index between 1.4~GHz and 5~GHz is flatter than $-0.5$, 
i.e.\ $\alpha \ge -0.5$ with $S_\nu \propto \nu^{\alpha}$.

\item The total flux density of (the components of) each source $\ge 30$~mJy 
 at 5~GHz.

\item The total flux density of (the components of) each source $\ge 20$~mJy 
 at 8.4~GHz. 

\item For multiply-imaged systems, (a) the compact radio-core images have 
separations $\ge 300$ milli-arcseconds, and (b) for doubly-imaged systems 
the ratio of the flux densities of the fainter to
the brighter images is $\ge 0.1$.
\end{enumerate} 
The final CLASS statistical sample includes JVAS sources which essentially
form the bright tail with their flux densities  $\ge 200$~mJy at 5~GHz.
However, not all JVAS sources are included in the final CLASS statistical 
sample because of the strict application of the above criteria (i) and (iv).

\begin{table*}
\caption{Multiply-imaged systems in the final CLASS statistical sample. 
The ratio of the flux densities of the images is given for the doubles only. 
Probable lens galaxy morphological-type identification is given using the 
codes, `e' for an early-type and `s' for a spiral. The references for the 
morphological type identifications are: (1) Browne et al.\ (1993); 
(2) Fassnacht \& Cohen (1998); (3) Impey et al.\ (1996); 
(4) Myers et al.\ (1995); (5) Sykes et al.\ (1998); 
(6) Augusto et al.\ (2001); (7) Rusin et al.\ (2001b).
\label{complete}}
\begin{tabular}{lllllllll}
\hline 
Source   &  Survey & Total flux &  Source   &  Lens & Maximum image 
 & Image flux-  & Number of  & Lensing \\
      &     &  density (mJy) & redshift $z_{s}$  &  redshift $z_{l}$ 
 & separation (arcsec) & density ratio  & images & galaxy type \\
\hline
0218+357  & JVAS  & 1480. &  0.96  & 0.68  & 0.334 & 0.26  & 2  & s$^1$ \\
0445+123  & CLASS & 50.  &  -     & -     & 1.33  & 0.14  & 2  & - \\
0631+519  & CLASS & 88.  &  -     & -     & 1.16  & 0.15  & 2  & - \\
0712+472  & CLASS & 30.  &  1.34  & 0.41  & 1.27  &  -     & 4   & e$^2$ \\
0850+054  & CLASS & 68.  &  -     & -     & 0.68  & 0.14  & 2  & - \\
1152+199  & CLASS & 76.  &  1.019 & 0.439 & 1.56  & 0.33  & 2  & - \\
1359+154  & CLASS & 66.  &  3.235 &  -    & 1.65  &  -   & 6    & - (3 Gs) \\
1422+231  & JVAS  & 548. &  3.62  & 0.34  & 1.28  &  -   & 4    & e$^3$ \\
1608+656  & CLASS  & 88. &  1.39  & 0.64  & 2.08  &  - & 4 & e (2 Gs)$^4$ \\
1933+503  & CLASS  & 63. &  2.62  & 0.755 & 1.17  &  -   & 4    & e$^5$? \\
2045+265  & CLASS  & 55. &  -   & 0.867  & 1.86  &  -   & 4  & - \\
2114+022   & JVAS  & 224. & - & 0.32/0.59 & 2.57 & 0.33 & 2? & e (2 Gs)$^6$\\
2319+051   & CLASS & 76.  &  - & 0.624/0.588 & 1.36  & 0.18 & 2 & e$^7$ \\
\hline
\end{tabular}
\end{table*}

The properties of the 13 multiply-imaged systems contained in the final CLASS
statistical sample are summarized in Table~1. Notice that two out of the four
multiply-imaged sources in the JVAS (statistical) sample used by Helbig et 
al.\ (1999), namely 0414+054 and 1030+074, are excluded because 0414+054 has
too steep a spectral index and 1030+074 has two images whose 
fainter-to-brighter image flux-density ratio is less than 0.1. 
Notice also that JVAS system 2114+022 which was excluded by Helbig et al.\ 
(1999) is included along with other multiple-galaxy lens systems.
The total flux densities, the lens redshifts (if measured), the source 
redshifts (see below for sources without measured redshifts), the image 
separations, the image multiplicities, and the lensing galaxy types (if 
determined), as given in Table~1, are all used (through the likelihood 
function defined in section~2.2) to constrain the parameters of the 
statistical lensing model. However, the measured image separations of 
1359+154, 1608+656, and 2114+022 are not used because the observed angular 
sizes are due to multiple galaxies within their critical radii and thus 
should not be used to constrain parameters pertaining to single galaxies. 
Notice particularly that lens modelling of these systems shows that the 
inferred critical radii of the secondary (and tertiary) galaxies are
comparable in size to those of the corresponding primary galaxies; the ratio
of the smallest to the largest critical radii ranges from 66\% to 81\%
assuming singular isothermal mass models for individual galaxies
[see Rusin et al.\ (2001a) for 1349+154, 
Koopmans \& Fassnacht (1999) for 1608+656, and Chae, Mao, \& Augusto (2001)
for 2114+022]. Furthermore, the inferred critical radii of the primary
galaxies for the multiple-galaxy lens systems are similar to those of the 
early-type lensing galaxies for the single-galaxy lens systems for which no 
comparably massive secondary galaxies are observed or required for lens 
modelling; specifically, the average critical radius of the primary galaxies 
for 1349+154, 1608+656, and 2114+022 is 1.34 arcsec 
while that for eight (presumably) early-type single-galaxy lens systems, 
namely 0445+123, 0631+519, 0712+472, 0850+054, 1152+199, 1422+231, 1933+503,
and 2319+051 (see below why 2045+265 is excluded), is 1.23 arcsec.
These results suggest that discarding the observed image separations for the
multiple-lens systems 1349+154, 1608+656, and 2114+022 would be an 
appropriate way of interpreting the data.
Likewise, the image multiplicities of the above three systems are not used 
to constrain mean mass ellipticity of galaxies because asymmetries in the 
total potentials of these systems are caused by combinations of multiple 
potentials. The image separation of 2045+265 is not used because of present
uncertainties in the details of the lensing scenario.
For the multiply-imaged sources without measured source redshifts,
we take $z_s = 2$ which is the mean source redshift for the multiply-imaged 
sources with measured source redshifts.

\subsection{Global properties of flat-spectrum radio sources}
As seen in section~2.1.2, the calculation of the lensing probability
(equations~29, 38, 39, or 40) for a source requires the knowledge of the
differential number--flux-density relation ($|dN_{z_s}(>S_{\nu})/dS_{\nu}|$) 
that the source follows and the redshift of the source ($z_s$). 
More specifically, the differential number--flux-density relation is 
required in the calculation of the magnification bias (equation~36). 
In particular, the differential number--flux-density relation 
needs to be known from the flux density of the source to
lower flux densities (formally to the zero flux density).
Furthermore, the differential number--flux-density relation needs to be known
as a function of source redshift. However, systematic measurements have been
done only for a few dozens of CLASS sources (Marlow et al.\ 2000). 
For this reason, we are led to ignore any redshift 
dependence of the differential number--flux-density relation. 
We also ignore weak dependences of the differential number--flux-density
relation on the spectral index of the source for the sake of simplicity.

The final CLASS statistical sample is well-described at $\nu = 5$~GHz by 
$|dN(>S_5)/dS_5| \propto (S_5/S_5^0)^{-\eta}$ with $\eta =2.07\pm 0.02$ 
for $S_5^0 < S_5 \la 1$~Jy where $S_5^0 = 30$~mJy (McKean et al., in 
preparation). 
For flux densities lower than $S_5^0$, McKean et al.\ (in preparation) find
$\eta =1.97\pm 0.14$ based on a Very Large Array (VLA) mapping of selected
regions of the sky down to $S_5 = 5$~mJy. 
The differential number--flux-density relation can also be estimated from
model flat-spectrum radio luminosity functions found in the literature.
Dunlop \& Peacock (1990, hereafter DP90) derived flat-spectrum radio 
luminosity functions based on redshift measurements of hundreds of flat 
spectrum radio sources with $S_{2.7} > 100$~mJy.
Specifically, DP90 presented five free-form radio luminosity functions 
and two parametric radio luminosity functions (i.e.\ a pure 
luminosity-evolution model and a model with both luminosity evolution and 
density evolution). 
Waddington et al.\ (2001) tested the DP90 models against the redshift
data for a Leiden Berkeley Deep Survey (LBDS) Hercules sample 
with $S_{1.4} > 1$~mJy.  Waddington et al.\ (2001) find that the measured
redshifts for the LBDS Hercules sample are consistent with two of the DP90
models, namely free-form model number 4 (FF-4) and free-form model number 5
(FF-5). {Fig.}~3 displays the measurements by McKean et al.\ (in preparation)
and the predictions by the DP90 FF-4 and FF-5 models for the differential
number--flux-density relation for flat-spectrum radio sources for $S_5 > 1$ 
mJy. As shown in {Fig.}~3, for flux densities below 30~mJy the measurements by
McKean et al.\ (in preparation) are roughly consistent with the extrapolation 
by the DP90 FF-5 model but inconsistent with the DP90 FF-4 model.
\begin{figure}
\begin{center}
\setlength{\unitlength}{1cm}
\begin{picture}(11,11)(0,0)
\put(-0.8,-0.8){\includegraphics{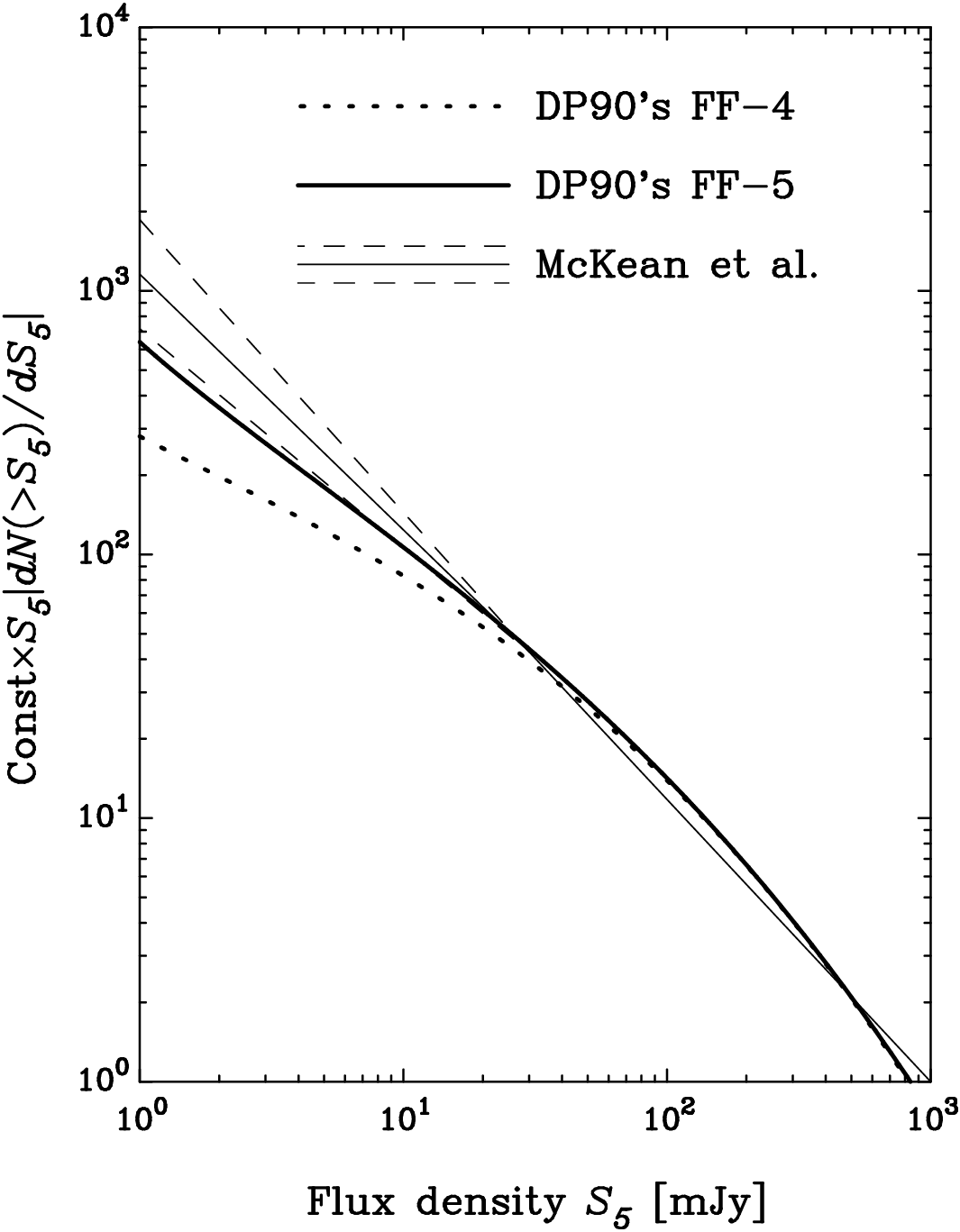}}
\end{picture}
\caption{
Differential number--flux-density relation $|dN(>S_5)/dS_5|$
for flat-spectrum radio sources. The measurement by McKean et al.\ 
(in preparation) is compared with the predictions by DP90's FF-4 and FF-5 
models (see section~3.2). 
}
\label{}
\end{center}
\end{figure}

Independent pieces of available evidence that can be used to 
estimate the redshift distribution of the CLASS sources are concordant. 
Based on spectroscopic observations of a representative subsample
of 42 CLASS sources with flux densities from $S_5=25$-50~mJy, 
Marlow et al.\ (2000) obtained a mean redshift $\langle z \rangle = 1.27$
with a dispersion of 0.95, at a completeness level of 64\%. 
Since the DP90 FF-4 and FF-5 models are consistent with the LBDS redshift data
with $S_{5} \ga 1$~mJy (see above), we are allowed to use the DP90 FF-4
and FF-5 models to infer mean redshifts for flat-spectrum sources with
$S_{5} \ga 1$~mJy.
\begin{figure}
\begin{center}
\setlength{\unitlength}{1cm}
\begin{picture}(8,8)(0,0)
\put(-3.,9.2){\includegraphics{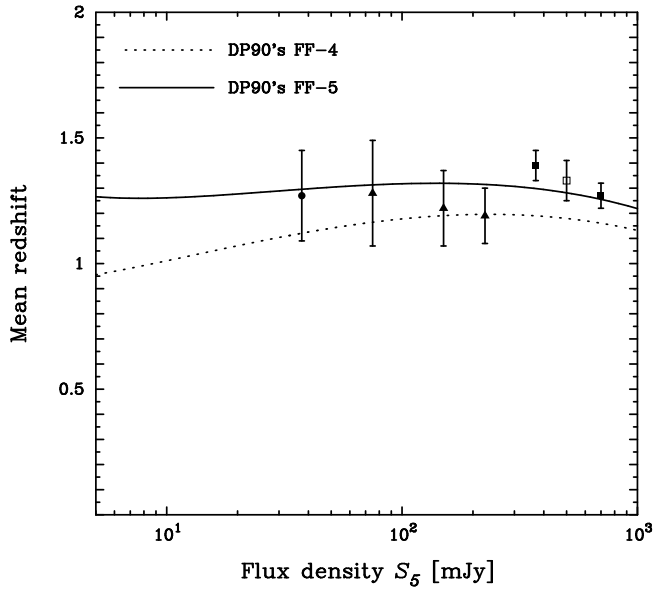}}
\end{picture}
\caption{
Mean redshifts of flat-spectrum radio sources as a function of flux density.
The measured mean and its $1\sigma$ error from various observations are
shown using a filled dot (CLASS subsample; Marlow et al.\ 2000), 
filled triangles (Falco, Kochanek, \& Mu\~{n}oz 1998), filled squares 
(Parkes quarter-Jansky sample; Jackson et al.\ 2002), and an open square
(the second Caltech-Jodrell Bank sample; Henstock et al.\ 1997).
The predictions by DP90's FF-4 and FF-5 models (see section~3.2) are 
respectively shown using dotted and solid lines.
}
\label{}
\end{center}
\end{figure}
{Fig.}~4 shows the mean redshift for flat-spectrum sources 
as a function of flux density predicted by the DP90 FF-4 and FF-5 models. 
{Fig.}~4 also shows results from various existing redshift observations of
flat-spectrum radio sources, including the Marlow et al.\ (2000) result 
and the results by Falco, Kochanek, \& Mu\~{n}oz (1998), 
Henstock et al.\ (1997), and Jackson et al.\ (2002). 
{From} {Fig.}~4 we see that the prediction by the DP90 FF-5 model agrees 
well with the independent data while the DP90 FF-4 model is marginally 
inconsistent with the data, similarly to the case for the differential 
number--flux-density relation (see above). This agreement increases 
our confidence in the Marlow et al.\ (2000) measured
mean redshift for the CLASS sample. Therefore, we adopt the measurement
by Marlow et al.\ (2000) for a mean redshift for the CLASS sources; i.e., 
we take a mean $\bar{z}  = 1.27 \pm 0.18$ where the estimated error is 
from the measured dispersion 0.95. 

Given that there are only 26 unbiased CLASS sources that have measured 
spectroscopic redshifts (Marlow et al.\ 2000), the distribution of the 
redshifts of the CLASS sources is poorly known. However, since 
mean redshifts for flat-spectrum sources are nearly the same for various
samples regardless of their flux density ranges, as shown in {Fig.}~4,
we may combine the samples assuming that the redshift distributions of the
samples are similar. The combined sample comprising the samples shown in
{Fig.}~4 has 747 flat-spectrum source redshifts with $S_5 \ga 30$~mJy.
A histogram of the 747 redshifts is shown in {Fig.}~5. Figure~5 also 
shows redshift distributions predicted by the DP90 FF-4 and FF-5 models
as well as a Gaussian model whose peak redshift and width are determined 
from the mean and dispersion measured by Marlow et al.\ (2000). From {Fig.}~5 
we see that the Gaussian model adequately describes the redshift distribution
for the combined sample. Therefore, we adopt the Gaussian model to approximate
the redshift distribution of the CLASS sources. 
\begin{figure}
\begin{center}
\setlength{\unitlength}{1cm}
\begin{picture}(7,7)(0,0)
\put(-2.2,8.1){\includegraphics{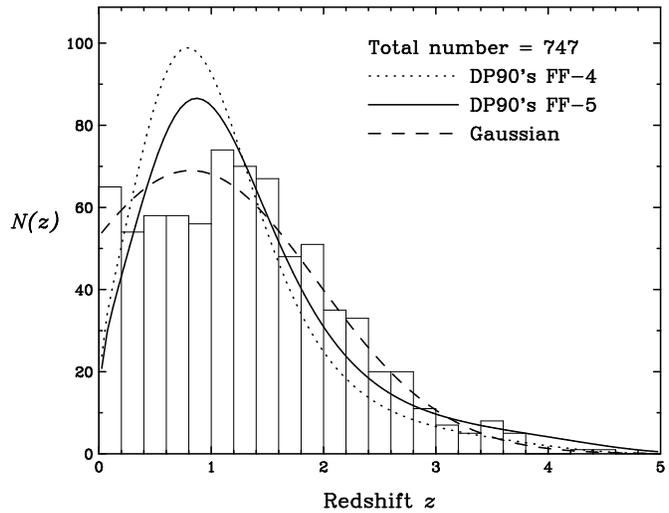}}
\end{picture}
\caption{
Redshift distribution of flat-spectrum sources. 
Redshift distributions predicted by the DP90 FF-4 and FF-5 models given
for $S_5 = 50$~mJy and a Gaussian model distribution (see section~3.2) are
overplotted on a histogram of 747 flat-spectrum source redshifts from various
observations (Henstock et al.\ 1997; Falco, Kochanek, \& Mu\~{n}oz 1998; 
Marlow et al.\ 2000; Jackson et al.\ 2002). 
}
\label{}
\end{center}
\end{figure}

\subsection{Galaxy luminosity functions per morphological type and their 
evolution since $z \sim 1$}
Use of reliable galaxy luminosity functions (LFs) is vital in analyses of 
statistical lensing. For example, use of an 
underestimated (overestimated) comoving number density of galaxies would
mislead us to overestimate (underestimate) a dark energy density.
In particular, given that galaxy populations of different morphological 
types contribute to multiple imaging in distinctively different ways 
(see section~2.1.3), it is required that type-specific galaxy LFs
 are reliably determined. For this reason, in this subsection
 we thoroughly review recent observational results on galaxy
LFs and obtain estimates of the early-type and the late-type 
LFs based on the data from the recent literature. 
We consider first the estimation of the local type-specific LFs
and then review recent observational studies on the evolution of early-type
galaxies since $z \sim 1$. 

We consider two approaches for estimating the local type-specific 
LFs. The first approach is to estimate the local total 
LF and decompose it into an early-type LF and a late-type
LF. The second approach is to simply adopt type-specific 
LFs measured in large-scale galaxy surveys.
The advantage of the first approach is that observational
information on the total LFs is abundant and now appears to 
produce a converging result on the total LF,
owing to more recent large galaxy redshift surveys particularly including 
the Sloan Digital Sky Survey (SDSS) and the Two Degree Field Galaxy Redshift 
Survey (2dFGRS). However, inference of the type-specific LFs
from the total LF relies on an independent knowledge of
{\it relative} type-specific LFs. By relative type-specific 
LFs, we refer to the relative magnitudes of 
the characteristic number densities of
the type-specific LFs. Thus, for the determination of the
relative type-specific LFs, the sample size of the galaxies
used is less relevant but the accuracy of the galaxy classification and the
photometry are important. Direct use of the measured type-specific 
LFs from large-scale galaxy surveys in which galaxies were classified
(the second approach) suffers from a relatively small number of independent 
measurements and uncertainties in classifications of large numbers of 
galaxies in each measurement.

We now consider estimating the total LF taking into 
account recent data from the literature. Observational results from various 
past galaxy surveys had systematic differences among 
the estimates of the local total LF. 
Not only were the measured values of the three Schechter parameters
$M_{*,0}$, $\alpha$ and $n_{*,0}$ significantly different 
among the various observational results, 
but also the estimates of the integrated luminosity density given by 
\begin{equation}
j_{0} = \int_0^{\infty} dL L \frac{dn(L,z=0)}{dL} = 
n_{*,0} L_{*,0} \Gamma(\alpha+2)
\end{equation}
differed by up to a factor of two.\footnote{
Differences in the measurements of each Schechter parameter are in part due to
degeneracies among the three Schechter parameters in fitting the data. 
However, the calculated quantity given by equation~(45) would not be 
significantly affected by such degeneracies.}
The past results indicate that, aside from any measurement errors 
(e.g.\ photometric errors, missing or misclassification of galaxies),
the various determinations of the total LF suffered from 
large-scale structures in the (local) Universe. 
The effects of large-scale structures 
on the determinations of the total LF
appear to have caused overall differences between measured LFs based on wide 
and shallow/sparse surveys (e.g.\ Loveday et al.\ 1992; Marzke et al.\ 1998, 
hereafter M98) and those based on narrow and deep surveys 
(e.g.\ Lilly et al.\ 1995; Ellis et al.\ 1996; Geller et al.\
 1997; Zucca et al.\ 1997), in the sense that normalisations of the 
LF for shallow samples are significantly lower than those for deeper 
samples. Systematic differences in the number counts of galaxies between 
those towards the northern Galactic cap and  
the southern Galactic cap also provide evidence of 
large-scale structures (Yasuda et al.\ 2001; Norberg et al.\ 2001). 
Thus, past determinations of the total LF depended, to a varying
extent, on the coordinates defining the sky patches observed in the surveys
as well as the angular areas and depths of the surveys. 

However, several recent galaxy surveys contain large numbers of 
galaxies because of the large areas and/or great depths of the
surveys. In addition, for a few of the recent galaxy surveys, 
CCD-based images of galaxies were used to measure the galaxy LFs.
More importantly, the results from several reliable observations appear to be
concordant. We consider the 2dFGRS (Cross et al.\ 2001, Norberg et al.\ 2001),
the SDSS (Blanton et al.\ 2001, Yasuda et al.\ 2001), the Century 
(Brown et al.\ 2001), the ESP (ESO Slice Project; Zucca et al.\ 1997), 
and the LCRS (Las Campanas Redshift Survey; Lin et al.\ 1996) surveys. 
The 2dFGRS (Cross et al.\ 2001, Norberg et al.\ 2001), the SDSS (Blanton et 
al.\ 2001, Yasuda et al.\ 2001), and the LCRS (Lin et al.\ 1996) observations
contain large numbers of galaxies (over 10,000 galaxies each).  
The SDSS (Blanton et al.\ 2001; Yasuda et al.\ 2001), the Century
(Brown et al.\ 2001), and the LCRS (Lin et al.\ 1996) observations use 
CCD-based photometry. Furthermore, the SDSS (Blanton et al.\ 2001; 
Yasuda et al.\ 2001) and the Century
(Brown et al.\ 2001) observations use multiple passbands.
All of the above recent surveys except for the
LCRS survey (Lin et al.\ 1996) are based on passbands similar to the 
Johnson-Morgan $B$ passband in wavelength coverage. The measured values of 
the Schechter parameters and the luminosity density (equation~45) 
from these recent surveys are summarized in Table~2. Notice that 
the characteristic magnitudes for all the above surveys except for the LCRS
are quoted in the $b_J$ photometric system using the relevant photometric
conversions quoted in the table.
\begin{table*}
\caption{Summary of recent measurements of the local total luminosity function
 (1. Cross et al.\ 2001; 2. Norberg et al.\ 2001; 3. Blanton et al.\ 
 2001; 4. Yasuda et al.\ 2001; 5. Brown et al.\ 2001;
  6. Zucca et al.\ 1997; 7. Lin et al.\ 1996)   }
\begin{tabular}{ccllll} \hline
 \# & Survey  &  $M_{*,0}-5\log_{10}h$  &  $\alpha$ & $n_{*,0}$  & $j_{0}$  \\
  &   &  &  & ($10^{-2}$ $h^3$ Mpc$^{-3}$) & 
 ($10^8$ $h$ $L_{\odot}$ Mpc$^{-3}$)   \\ 
 \hline
 1 & 2dFGRS   & $-19.75\pm 0.05$ ($b_J$)  & $-1.09\pm 0.03$  & $2.02\pm 0.02$ 
 & $2.24\pm 0.12$   \\
 2 &  2dFGRS   & $-19.66\pm 0.07$ ($b_J$)   & $-1.21\pm 0.02$ & $1.68\pm 0.08$
 & $1.90\pm 0.18$    \\
 3 &  SDSS     &  $-19.71\pm 0.05$ ($b_J$)$^a$ & $-1.26\pm 0.05$  & 
 $2.06\pm 0.23$ & $2.57\pm 0.34$    \\
 4 &  SDSS   &  $-19.71\pm 0.05$ ($b_J$)$^a$  &  $-1.26\pm 0.05$ & 
 $1.81\pm 0.16$  & $2.26\pm 0.26$     \\
 5 &  Century   & $-19.5 \pm 0.09$  ($b_J$)$^b$  & $-1.07\pm 0.09$ & 
 $2.0\pm 0.3$  &  $1.74\pm 0.32$   \\
 6 &  ESP   &  $-19.61^{+0.06}_{-0.08}$ ($b_J$)   &  $-1.22^{+0.06}_{-0.07}$
  &  $2.0\pm 0.4$   & $2.19\pm 0.48$    \\
 7 &  LCRS  & $-20.29\pm 0.02$ ($R_{KC}$) & $-0.70\pm 0.05$  & $1.9\pm 0.1$  
   &  $1.4\pm 0.1$    \\ \hline
  & Our estimate  &  $-19.69\pm 0.04$ ($b_J$)   & $-1.22\pm 0.02$  &
  $1.71\pm 0.07$   &  $2.02\pm 0.15$    \\ \hline
\end{tabular}

$^a$ The SDSS $g^*$ magnitude was converted to the $b_J$ magnitude using
 $b_J = g^* + 0.12 + 0.16 (B-V)$ and $B-V = 0.94$ (Norberg et al.\ 2001)

$^b$ The Century $V$ magnitude was converted to the $b_J$ magnitude using
    $b_J - R = 1.2$ and $V-R = 0.53$ (Brown et al.\ 2001).

\end{table*}

The Norberg et al.\ (2001) result was obtained by normalising their 
LF to the combined galaxy number counts from the 2dFGRS southern and
northern Galactic caps (over 110,500 galaxies). Likewise, the Yasuda et al.\
(2001) result was obtained by correcting the LF calculated by
Blanton et al.\ (2001) to match the number counts of galaxies in SDSS 
($\sim 900,000$ galaxies). The Norberg et al.\ (2001) result 
included modelling and correcting for galaxy evolutionary effects, and 
the SDSS results were derived using Petrosian magnitudes of galaxies (see
Blanton et al.\ 2001), which overcome dependences of isophotal magnitudes
on the surface-brightness distribution of galaxies. The Norberg et al.\ 
(2001) and the Yasuda et al.\ (2001) results are independent of 
and consistent with each other and probably represent present 
best estimates of the local total LF. 
The results of the Century (Brown et al.\ 2001) and
the ESP (Zucca et al.\ 1997) surveys are also consistent with the Norberg
et al.\ (2001) and the Yasuda et al.\ (2001) results; 
however, their statistical errors are larger. 
The result of the LCRS (Lin et al.\ 1996), which is based
on an $R$ passband, appears to be significantly different from  those of the
other surveys. We consider the LCRS (Lin et al.\ 
1996) because of the large comoving volume covered by the survey and the
fact that it is one of the few CCD-based imaging surveys.
While the characteristic magnitude and normalisation of the LF
depend on the wavelength coverage of the passband used, recent determinations
of LFs at multiple passbands based on CCD photometry
did not show any evidence for dependence of the faint-end slope on passbands 
of optical wavelengths (Blanton et al.\ 2001; Brown et al.\ 2001). 
Thus, the significantly shallower faint-end slope obtained from 
the LCRS (Lin et al.\ 1996) compared with those from the other surveys given 
in Table~2 might appear to be a significant obstacle to a concordant
total LF. However, using the SDSS data, Blanton et al.\ 
(2001) illustrate dependences of the chosen isophotal limits on the derived 
LF parameters. In particular, Blanton et al.\ (2001) argue
that the result of Lin et al.\ (1996) was significantly affected by their 
chosen relatively bright isophotal limit. We use the Norberg et al.\ (2001)
result and the Yasuda et al.\ (2001) result to estimate
the Schechter parameters of the local total LF. Although the 
errors of the three Schechter parameters may be correlated, we ignore such
correlations and calculate weighted means of the values of the parameters 
obtained by Norberg et al.\ (2001) and Yasuda et al.\ (2001). This result is
our estimate of the local total LF given in 
Table~2.\footnote{We notice that our estimate of the Schechter parameters
is also in good agreement with just recently published galaxy counts in the 
Millennium Galaxy Catalogue (Liske et al.\ 2002).}
 
To extract type-specific LFs from our estimated total 
LF, we define the ratio of the integrated luminosity
density of a population of galaxies of type $g$ ($g=e,s$) to 
the total integrated luminosity density, i.e., 
\begin{equation}
\zeta^{(g)}_0 \equiv  \frac{j_0^{(g)}}{j_0} = 
\frac{n_{*,0}^{(g)}}{n_{*,0}} 10^{0.4(M_{*,0}-M_{*,0}^{(g)})} 
\frac{\Gamma(\alpha^{(g)}+2)}{\Gamma(\alpha+2)},
\end{equation}
where the last equality follows from equation~(45). Notice that the 
fractional luminosity density given by equation (46) is unaffected either
by an error in the absolute photometry or 
large-scale galaxy-number-density fluctuations. The accuracy in the
measurement of the fractional luminosity density in a survey
is limited mainly by errors in relative photometry and errors in 
the galaxy classification. It follows from equation~(46) that 
the type-specific normalisation $n_{*,0}^{(g)}$ is given by
\begin{equation}
n_{*,0}^{(g)} = \zeta^{(g)}_0 n_{*,0} 10^{0.4(M_{*,0}^{(g)}-M_{*,0})}
  \frac{\Gamma(\alpha+2)}{\Gamma(\alpha^{(g)}+2)}.
\end{equation}
Given an estimate of the total LF, independent 
estimates of parameters $\alpha^{(g)}$ and $M_{*,0}^{(g)}$ along with the
the fractional luminosity density $\zeta^{(g)}_0$ will allow us
to estimate the type-specific normalization $n_{*,0}^{(g)}$ using 
equation~(47).

We now consider direct measurements of the early-type Schechter parameters 
from recent galaxy surveys and derive corrected early-type Schechter
normalisations using equation~(47) from the directly measured early-type
Schechter parameters and our estimate of the total LF
given in Table~2.
Direct measurements of the early-type LF from galaxy surveys
are relatively few and have relatively larger uncertainties mainly because of
the challenging observational task of accurately classifying large numbers of
galaxies. The observational techniques of classifying galaxy types used 
in various galaxy surveys are based on morphological appearances,
surface brightness profiles, prototype spectra, 
 spectral energy distributions, and/or colours of galaxies.
Identifying early-type galaxies based on morphological
appearances or surface brightness profiles is probably the most 
reliable method, since the morphological appearances and 
the light profiles of spirals, which constitute the most numerous population, 
are well distinguished from those of early-types. 
We consider the results from the Second Southern Sky Redshift Survey (SSRS2;
M98) and the 2dFGRS (Folkes et al.\ 1999; 
Madgwick et al.\ 2001).\footnote{An estimate of the Schechter parameters 
per galactic type is not yet reported from the SDSS, although some
classifications of galaxies have been considered (Blanton et al.\ 2001; 
Bernardi et al.\ 2003). In particular, we note that the sample of
early-type galaxies used by Bernardi et al.\ (2003) in their study of
early-type galaxies were selected too
stringently (for example, luminosities of galaxies in the sample spread over 
only 4 magnitudes in the brightest range) and thus not suitable to derive a 
`fair' LF of early-type galaxies.}
The SSRS2 (M98) is a wide-angle, 
shallow ($z \le 0.05$) survey of 5404 galaxies 
in the northern and southern Galactic caps that are brighter than
$m_B = 15.5$. The relatively bright magnitude limit used for the SSRS2
allowed M98 to classify the galaxies in their sample 
through the visual inspection of morphological appearances with more than 
99\% completeness. M98's determination of the early-type
LF probably represents the best estimate of the local early-type LF. 
However, as they discuss, the M98
sample would not provide a fair number density of galaxies because 
the relatively small volume covered by their sample suffers from large-scale
density fluctuations. In this respect, the M98 result would
only be valuable as a relative early-type LF. 
M98's measured values of the local early-type 
LF parameters can be found in Table~3. 
\begin{table*}
\caption{Summary of estimates of the local early-type luminosity function
 (1. M98; 2. Folkes et al.\ 1999; 3. Madgwick et al.\ 2001).
The normalisations in the original survey results are corrected using 
equation~(47) and the total luminosity function (Table~2).}
\begin{tabular}{cclllll} \hline
\# & Survey &  $ M_{*,0}^{(e)}-5\log_{10}h$  & $\alpha^{(e)}$  
& $\zeta_0^{(e)}$  & $n_{*,0}^{(e)}$ & $j_{0}^{(e)}$  \\  
  &  &   &  &   & ($10^{-2}$ $h^3$ Mpc$^{-3}$) &
  ($10^8$ $h$ $L_{\odot}$ Mpc$^{-3}$)   \\  \hline
 1& SSRS2 & $-19.63^{+0.10}_{-0.11}$ ($b_J$)$^a$ & $-1.00\pm 0.09$ 
 & $0.30\pm 0.08$  & $0.44\pm 0.08$ & $0.41\pm 0.09$   \\
 & normalisation corrected &   &   &   & $0.64\pm 0.19$   & $0.60\pm 0.19$  \\
 2 & 2dFGRS & $-19.61\pm 0.09$ ($b_J$)  & $-0.74\pm 0.11$ 
 & $0.34\pm 0.06$  & $0.90\pm 0.09$ & $0.75\pm 0.10$  \\
 & normalisation corrected &  &  &  & $0.82\pm 0.17$   & $0.68\pm 0.15$ \\
 3 & 2dFGRS   & $-19.59\pm 0.05$ ($b_J$)  & $-0.54\pm 0.02$  & $0.40\pm 0.04$
 & $0.99\pm 0.05$  & $0.79\pm 0.05$    \\  
 & normalisation corrected &  &  &  & $1.01\pm 0.13$  &  $0.81\pm 0.11$  \\
 \hline
\end{tabular}

$^a$ The $B(0)$ magnitude of SSRS2 was converted to $b_J$ magnitude 
 using $b_J = B - 0.28 (B-V)$ (Blair \& Gilmore 1982; Norberg et al.\ 2001)
 and assuming $B(0) = B$ (Alonso et al.\ 1993) and mean $(B-V)=0.94$ 
 (Norberg et al.\ 2001).
\end{table*}

Apart from the M98 study, recent
observational studies on galaxy LFs per type did not
classify galaxies through morphological appearances but used 
classification techniques that are based on photometric and spectroscopic 
information on galaxies (e.g.\ Folkes et al.\ 1999; Blanton et al.\ 2001; 
Fried et al.\ 2001; Madgwick et al.\ 2001; Bernardi et al.\ 2003). 
We consider the results by Folkes et al.\ (1999) and Madgwick et al.\ (2001) 
which are based on the 2dFGRS data.
Folkes et al.\ (1999) and Madgwick et al.\ (2001) have employed
the so-called ``principal component analysis (PCA)'' of galaxy spectra, 
which essentially maximally quantifies differences between the spectra of 
galaxies. Kochanek, Pahre, \& Falco (2000) have recently argued that 
the physical properties of apertures (particularly their sizes) 
used for spectroscopic observations in various galaxy redshift surveys 
are significant sources of systematic errors in galaxy classifications. 
More generally, those techniques will give correct results only up to
the accuracy of the assumed relation between the morphological types and 
(the parameters used to describe) the spectra of galaxies. Indeed,
applying spectrum-based classification techniques to a sample of 4000
2dFGRS galaxies whose morphological types are visually known, Madgwick (2003)
finds that the rates of successfully classifying early-type and late-type
galaxies by spectrum-based classification techniques range from 
$\sim 60$\%-80\%. Nonetheless, the results by Folkes et al.\ (1999) 
and Madgwick et al.\ (2001) are some of the few relatively more reliable 
results that are available. The values of the Schechter parameters for the 
local early-type galaxy population obtained by  Folkes et al.\ (1999) and 
Madgwick et al.\ (2001) can be found in Table~3.
In Table~3, the corrected early-type normalisations using equation~(47) can
also be found next to each survey result. While the corrected normalization
is higher for the SSRS2 result, it is essentially not changed either for the 
Folkes et al.\ (1999) result or the Madgwick et al.\ (2001) result.
This implies that the total luminosity densities measured by Folkes et al.\ 
(1999) and Madgwick et al.\ (2001) are consistent with our estimate while
that measured by M98 is lower than our estimate. 
In section~4, we will use for our analyses the corrected SSRS2
(M98) early-type LF and the early-type 
LF directly measured by Madgwick et al.\ (2001). 

Since the total luminosity density is dominated by late-type galaxies, we 
assume that the faint-end slope and the characteristic absolute magnitude for 
late-type galaxies are the same as those for all galaxies, namely we take 
$\alpha^{(s)} = -1.22 \pm 0.02$ and $M_{*,0}^{(s)} = -19.69\pm 0.04$ ($b_J$). 
Under this prescription, the normalisation for the late-type population is 
the same as the ratio of the luminosity density of the late-type  population 
to the total luminosity density. Hence the inferred normalisations for the 
late-type population are $n_{*,0}^{(s)} = 1.20\pm 0.11$ $10^{-2}$ $h^3$ 
Mpc$^{-3}$ for the M98 type-specific LFs, 
$n_{*,0}^{(s)} = 1.13\pm 0.09$ $h^3$ Mpc$^{-3}$ for
the Folkes et al.\ (1999) type-specific LFs, and 
$n_{*,0}^{(s)} = 1.03\pm 0.08$ $10^{-2}$ $h^3$ Mpc$^{-3}$ 
for the Madgwick et al.\ (2001) type-specific LFs.

Finally, it is required that galaxy evolution (of relevance to lensing) 
be incorporated in the analyses of statistical lensing because 
the lensing probability is sensitive to the redshift of
the potential lensing galaxy (Turner, Ostriker \& Gott 1984). Since 
the lensing rate is dominated by early-type galaxies, the most relevant
issue of galaxy evolution in statistical lensing is whether there is 
any significant change in the comoving number density of early-type galaxies 
at most likely lens redshifts (i.e.\ $z \sim 0.6$) compared with the present
epoch as would be the case in certain rapid-evolution models
(e.g.\ Baugh, Cole, \& Frenk 1996). 
There exist in the literature two qualitatively different kinds of
observational results on the formation and evolution of 
early-type galaxies, each of which supports an extreme view of the 
formation and evolution of early-type galaxies; namely, an early formation by
the monolithic collapse of protogalactic gases followed by passive evolution
(Eggen, Lynden-Bell \& Sandage 1962) or a continuous formation through
merging of subunits in a hierarchical structure-formation model (Larson 1974;
White \& Rees 1978). Galaxy count results that are consistent with the
early formation and passive evolution of early-type galaxies appear to be
favoured (Im et al.\ 1996, 1999, 2002; Schade et al.\ 1999; 
Totani \& Yoshii 1998; Lilly et al.\ 1995) while some results would support
a rapid evolution of early-type galaxies since an intermediate to high 
redshift (Fried et al.\ 2001; Kauffmann, Charlot \& White 1996). Results 
from the analysis by the fundamental plane of a sample of 
gravitational-lens early-type galaxies (Kochanek et al.\ 2000)
would not be consistent with a rapid evolution of early-type galaxies but
would be consistent with the early formation and passive evolution hypothesis.
A further review of the current status on the formation and evolution of 
early-type galaxies can be found in Peebles (2002), who argues for
the early formation and passive evolution of early-type galaxies based on
numerous lines of observational studies such as the colour-magnitude relation
and the fundamental plane. Following Peebles (2002) and the authors 
mentioned above who argue for passive evolution of early-type galaxies,
it will be assumed in this study that the local early-type LF
has not evolved in the sense that the characteristic comoving
number density and the faint-end slope of the early-type LF are unchanged in
the lookback times.

\section{RESULTS}
The results on cosmological parameters and global properties of galaxies
that we derive in this section are under the following two key assumptions:
\begin{itemize}
\item A galaxy as a gravitational lens is assumed to be well-approximated by
a singular isothermal ellipsoid mass model. Furthermore, dynamical
normalisations of lenses are only considered for the oblate and the prolate
cases.

\item The early-type comoving characteristic number density and faint-end
slope are assumed to be unchanged from $z \sim 1$ to the present epoch. 
\end{itemize} 
We emphasize that the above assumptions are valid at least as a first-order 
approximation according to the majority of current evidence (section~2.1; 
section~3.3). The parameters of our statistical lensing model are summarized
in section~2.1.3. Our goals are to constrain cosmological parameters 
and the early-type characteristic velocity dispersion (and the late-type
characteristic velocity dispersion) and the mean projected mass ellipticity 
of (early-type) galaxies from the observed properties of statistical lensing
(given in section~3.1). To meet our goals, we need to fix the rest of the
parameters because there exist degeneracies in the total parameter space
with the observed properties of statistical lensing. Recent advanced 
observations have determined the fixed parameters with relatively small
uncertainties, as given in section~3.2, section~3.3, and below. Nonetheless,
uncertainties in some parameters remain possible sources of systematic errors
in the derived results and are discussed in sections~4.3 and 5.3. 

The fixed parameters are as follows: (1) both the early-type and the late-type
LF faint-end slopes and characteristic number densities as given in 
section~3.3 (note, however, that two alternative choices of the type-specific
LFs are used; see below); (2) the Faber-Jackson exponent 
$\gamma_{\rm FJ} = 4.0$ (Bernardi et al.\ 2001)
and the Tully-Fisher exponent $\gamma_{\mbox{\scriptsize TF}} = 2.9$
(Tully \& Pierce 2000) (these parameters affect relatively weakly  
constraints on galaxy velocity dispersions and consequently cosmological 
parameters within their present uncertainties);
(3) the parameters describing the redshift distribution and 
the differential number--flux-density relation for the CLASS sources as
 given in section~3.2 (systematic uncertainties arising from the uncertainties
 in these parameters will be estimated separately).
As simplifications of our analyses, we use a common mean projected mass 
ellipticity $\bar{\epsilon}$ and a common relative frequency of the 
oblatenesses ${\mathcal P}_{\mbox{\scriptsize obl}}$ both for the early-type
and the late-type populations. The mean ellipticity and the intrinsic shape 
distribution of late-type galaxies have little effect on our derived results
(other than the dependence of the derived late-type characteristic velocity 
dispersion on these parameters), so we simplify our analyses by using the 
above prescriptions. The parameters that are allowed to be free are as 
follows: (1) the cosmological parameters; matter density $\Omega_{\rm m}$,
density of a classical cosmological constant $\Omega_{\Lambda}$ or 
dark energy density $\Omega_x$ with its constant equation of state $w$; 
(2) the early-type and the late-type characteristic velocity dispersions 
$\sigma_{*}^{(e)}$ and $\sigma_{*}^{(s)}$; (3) the mean projected mass 
ellipticity of (early-type) galaxies $\bar{\epsilon}$ and the relative 
frequency of the oblatenesses ${\mathcal P}_{\mbox{\scriptsize obl}}$ in the
intrinsic shape distribution of (early-type) galaxies.
We emphasize that it is not only justified but also very sensible to treat the
characteristic velocity dispersions (in particular $\sigma_{*}^{(e)}$) as
free parameters and determine them from the observed image separations 
in the final statistical sample. This is because the observed radio-image
separations are very accurate and as can be seen from equation~(35), 
image separation has a strong sensitivity on velocity dispersion but little
on cosmological parameters in particular within the ranges to be considered
below.

While we allow the early-type and the late-type characteristic velocity
dispersions to be free, a characteristic central stellar velocity dispersion
$\sigma_{{\rm c}*,0}^{(e)}$ (at the present epoch) for the early-type 
population can be obtained through a measured Faber-Jackson 
relation from the early-type characteristic absolute magnitude,
and a characteristic maximum rotation speed $v_{{\rm max}*,0}^{(s)}$ 
(at the present epoch) for the late-type population 
can also be obtained through a measured Tully-Fisher relation 
from the late-type characteristic absolute magnitude.
The early-type characteristic central stellar velocity dispersion and the 
late-type characteristic maximum rotation speed (obtained in the above ways)
can then be used to estimate the early-type and the late-type characteristic
velocity dispersions, although the conversions are model-dependent. 
Below we use these methods to obtain independent estimates of the early-type
and the late-type characteristic velocity dispersions. Such estimates can be
compared with the derived values from the statistical lensing model fitting.
Furthermore, we will use the estimated late-type characteristic velocity 
dispersion as a prior constraint for a model fitting in section~4.1.
This choice of model fitting is motivated because there is only one 
confirmed late-type lensing galaxy in the final CLASS statistical sample that
can be used to constrain the late-type characteristic velocity dispersion.

The local early-type characteristic absolute magnitude of 
$M_{*,0}^{(e)}(b_J) - 5\log_{10}h = -19.61$ along with $B=b_J+0.26$ 
(section~3.3) corresponds to a characteristic central stellar velocity 
dispersion of $\sigma_{{\rm c}*,0}^{(e)} = 192\pm 34$ km~s$^{-1}$ through
$-M_B = A + 9(\log_{10} \sigma_{\rm c} -2.3)$ with a mean $A = 19.5\pm 0.7$ 
for both ellipticals and S0 galaxies (expressed for $h=1$; 
de~Vaucouleurs \& Olson 1982). 
By fitting the observed light distributions and the observed velocity 
dispersion profiles of 37 elliptical galaxies using singular isothermal 
mass distributions, Kochanek (1994) finds that the central stellar velocity 
dispersion of an elliptical galaxy is a good measure of the velocity 
dispersion of a singular isothermal mass model. Following Kochanek (1994),
we thus estimate that $\sigma_{*,0}^{(e)} \approx \sigma_{{\rm c}*,0}^{(e)} =
192\pm 34$ km~s$^{-1}$. Alternatively,
using the `Faber-Jackson' relation for the dark matter velocity dispersion
(i.e., the $\sigma_{\rm DM}$-$L$ relation) shown in {Fig.}~4 of Kochanek 
(1994), we find that the above early-type characteristic absolute magnitude 
corresponds to an early-type characteristic velocity dispersion of 
$\sigma_{*,0}^{(e)} = 185^{+9}_{-7}$ km~s$^{-1}$. 

We adopt the following Tully-Fisher relation to obtain a characteristic
velocity dispersion $\sigma_{*,0}^{(s)}$ for the late-type population 
(Tully \& Pierce 2000):
\begin{equation}
-M^{\rm c}_B = 7.27 [\log_{10} (2v_{\rm max}^{(s)}) - 2.5] + 20.11,
\end{equation}
where $v_{\rm max}^{(s)}$ is the maximum rotation speed (i.e.\ for the
edge-on view) and $M^{\rm c}_B$ is the extinction-corrected absolute $B$
magnitude (i.e., the magnitude for the face-on view), which is given by
\begin{equation}
M_B^{\rm c} = M_B^{\rm ob} - 
\{1.57+2.75[\log_{10} (2v_{\rm max}^{(s)}) - 2.5]\} \log (a/b),
\end{equation}
where $M_B^{\rm ob}$ is the uncorrected absolute magnitude and
$a/b$ is the apparent major-to-minor axis ratio. From
 $M_{*,0}^{(s)} (b_J) -5 \log_{10} h = -19.69\pm 0.04$ along with 
$B=b_J+0.26$ (section~3.3) and $h=0.72\pm 0.08$ (Freedman et al.\ 2001),
we obtain the characteristic maximum rotation speed of 
$v_{{\rm max}*,0}^{(s)} = 189^{+17}_{-15}$ km~s$^{-1}$ 
using equations~(48) and (49) and taking a mean $a/b=2$. 
If we assume an SIS mass distribution for the halo of a late-type galaxy, 
this characteristic maximum rotation speed corresponds to 
the characteristic velocity dispersion of $\sigma_{*,0}^{(s)}
= v_{{\rm max}*,0}^{(s)}/\sqrt{2} = 134^{+12}_{-10.6}$ km~s$^{-1}$.

As discussed at length in section~3.3, while recent observational 
determinations of the total galaxy luminosity function give 
a converging result, the decomposition of the total luminosity function by 
morphological type is less certain. For this uncertainty, we do our analyses 
for the two choices of the type-specific galaxy LFs, i.e.\
the normalisation-corrected SSRS2 type-specific LFs and the 
2dFGRS type-specific LFs (see Table~3 and section~3.3).
The SSRS2 type-specific LFs\footnote{Hereafter, `SSRS2 
type-specific LFs' always refer to the normalisation-corrected SSRS2 
type-specific LFs.} 
will be our standard choice because their galaxies were classified through
visual identifications (see section~3.3), and 
the 2dFGRS type-specific LFs 
will be our alternative choice. Notice that the SSRS2 early-type LF has 
a significantly lower characteristic number density and a significantly
steeper faint-end slope compared with the 2dFGRS early-type LF
[i.e.\ $n_{*,0}^{(e)}/(10^{-2}$~$h^3$~Mpc$^{-3}$) = $0.64\pm 0.19$ 
vs.\ $0.99\pm 0.05$ and $\alpha = -1.00\pm 0.09$ vs.\ $-0.54\pm 0.02$].

\subsection{Constraints on cosmological parameters}
We first consider cosmological models with a classical cosmological constant
in which the cosmological parameters to be constrained are the matter density
$\Omega_{\rm m}$ and the cosmological constant density $\Omega_{\Lambda}$. 
In other words, we only consider a fixed constant equation of state for dark
energy, $w = -1$. Notice, however, that even if we allow $w$ along with 
$\Omega_{\rm m}$ and $\Omega_{\Lambda}$ to be varied simultaneously, the
constraints on the  $\Omega_{\rm m}$-$\Omega_{\Lambda}$ plane would change 
little because the best-fitting value of $w$ is $-1$ (with the prior 
constraint $w \ge -1$ imposed) at least for cosmological models that are near
the flat line $\Omega_{\rm m} + \Omega_{\Lambda} = 1$.
We then consider flat cosmological models with a dark energy whose equation of
state is assumed to be constant and satisfies the energy condition $w \ge -1$.
In this case, the cosmological parameters to be constrained are the matter
density $\Omega_{\rm m} (=1-\Omega_x)$ and the constant dark-energy equation
of state. Notice here that because of degeneracy on the 
$\Omega_{\rm m}$-$\Omega_{\Lambda}$ plane along a line that is nearly 
orthogonal to the flat line for a given value of $w$, it is useful to impose
the constraint $\Omega_{\rm m} + \Omega_{\Lambda} = 1$.

{Fig.}~6 shows confidence limits in the parameter plane spanned by 
$\Omega_{\rm m}$ and $\Omega_{\Lambda}$. At each grid point on the plane
the value of the $\chi^2$ (equation~44) is determined by minimizing the 
function over the nuisance parameters (i.e., all the other free parameters). 
The four contours respectively correspond to 68\%, 90\%, 95\%, 
and 99\% confidence limits for one parameter, namely
$\Delta\chi^2 = 1.0$, 2.71, 4.0, and 6.63 from the global minimum value of
the $\chi^2$, $\chi^2_{\rm min}$. Notice that the adopted method for 
estimating confidence limits on model parameters is the so-called (profile)
likelihood ratio method which is one of the widely practiced methods 
[see, e.g., Lupton (1993) and references therein]. 
This method gives more conservative (i.e.\ larger)
confidence intervals than the frequently practiced method in which nuisance
parameters are held fixed at their global maximum-likelihood values. 
The confidence intervals obtained by the likelihood ratio method also closely
resemble realistic confidence intervals obtained by projecting the 
n-dimensional confidence region onto the parameter subspace of interest.
However, while computation of the likelihood distribution in the n-dimensional
(in the present case 5-dimensional) parameter space is costly even for a 
moderately grained grid, distribution of likelihood in one or two-dimensional
parameter subspace can be computed easily using an efficient optimization
routine such as the downhill simplex method (see Press et al.\ 1992).
Here and throughout in this paper all the confidence limits on model 
parameters are based on the likelihood ratio method. However, we also
compute a likelihood distribution in a pseudo full-dimensional grid. By the
likelihood distribution in a pseudo full-dimensional grid we shall mean the
likelihood distribution in the subspace spanned only by the most crucial
parameters where each likelihood value is obtained by fixing the rest of the
parameters at their maximum-likelihood values. For the case under 
consideration, the most crucial three parameters are $\Omega_{\rm m}$, 
$\Omega_{\Lambda}$, and $\sigma_{*}^{(e)}$. The pseudo full-dimensional
likelihood distribution is available at the Jodrell Bank Observatory web page
for gravitational lenses and the 
CLASS.\footnote{http://www.jb.man.ac.uk/research/gravlens/numerical/numerical.html}
The web page also includes (profile)
likelihood distributions in parameter planes.

{Fig.}~6~(a) and (b) are the results
based on the SSRS2 type-specific LFs while {Fig.}~6~(c) and (d) are those
based on the 2dFGRS type-specific LFs. For the results of {Fig.}~6~(b) 
and (d), the late-type characteristic velocity dispersion is constrained
by the value obtained from the Tully-Fisher relation (see above). 
\begin{figure*}
\begin{center}
\setlength{\unitlength}{1cm}
\begin{picture}(18,18)(0,0)
\put(0.,-2.){\includegraphics{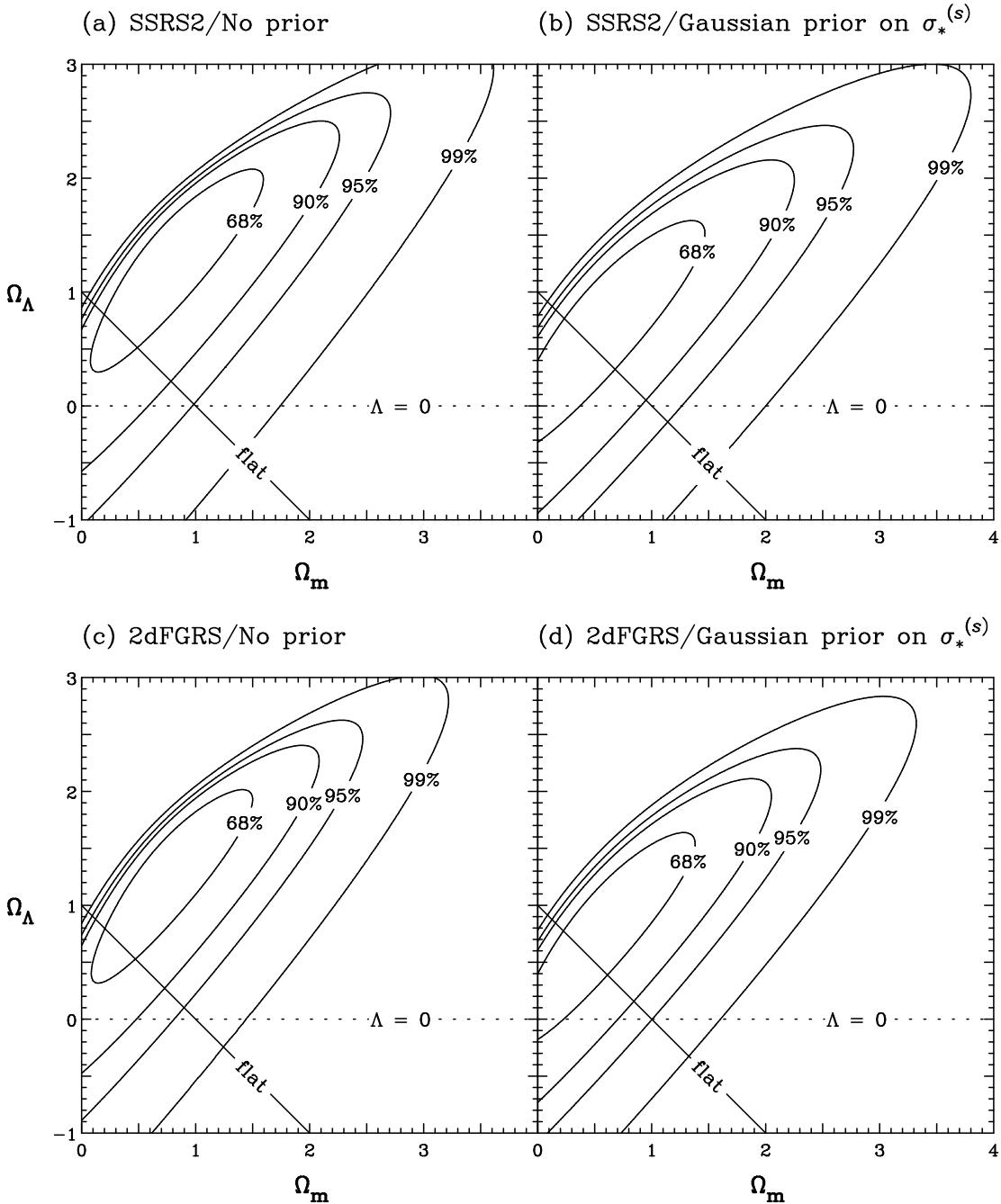}}
\end{picture}
\caption{
Likelihood regions in the $\Omega_{\rm m}$-$\Omega_{\Lambda}$ plane.
The four contours, respectively, correspond to the 68\%, 90\%, 95\%, and 99\%
confidence limits using the $\Delta\chi^2$ (or, likelihood ratio) statistic
for one parameter. The four illustrated models are as follows. Models (a) and
(b) are based on the SSRS2 type-specific LFs while models (c) and (d) are
based on the 2dFGRS type-specific LFs. For models (a) and (c), no prior
constraints are imposed on the nuisance parameters while for models (b) and
(d), $\sigma_{*,0}^{(s)} = 134_{-10.6}^{+12}$~km~s$^{-1}$ (see section~4) 
in conjunction with the dynamical normalisation for the equal frequencies of 
oblates and prolates is used as a prior constraint assuming 
$\sigma_{*}^{(s)}=\sigma_{*,0}^{(s)}$.
}
\label{}
\end{center}
\end{figure*}
\begin{figure*}
\begin{center}
\setlength{\unitlength}{1cm}
\begin{picture}(18,18)(0,0)
\put(0.,-2.){\includegraphics{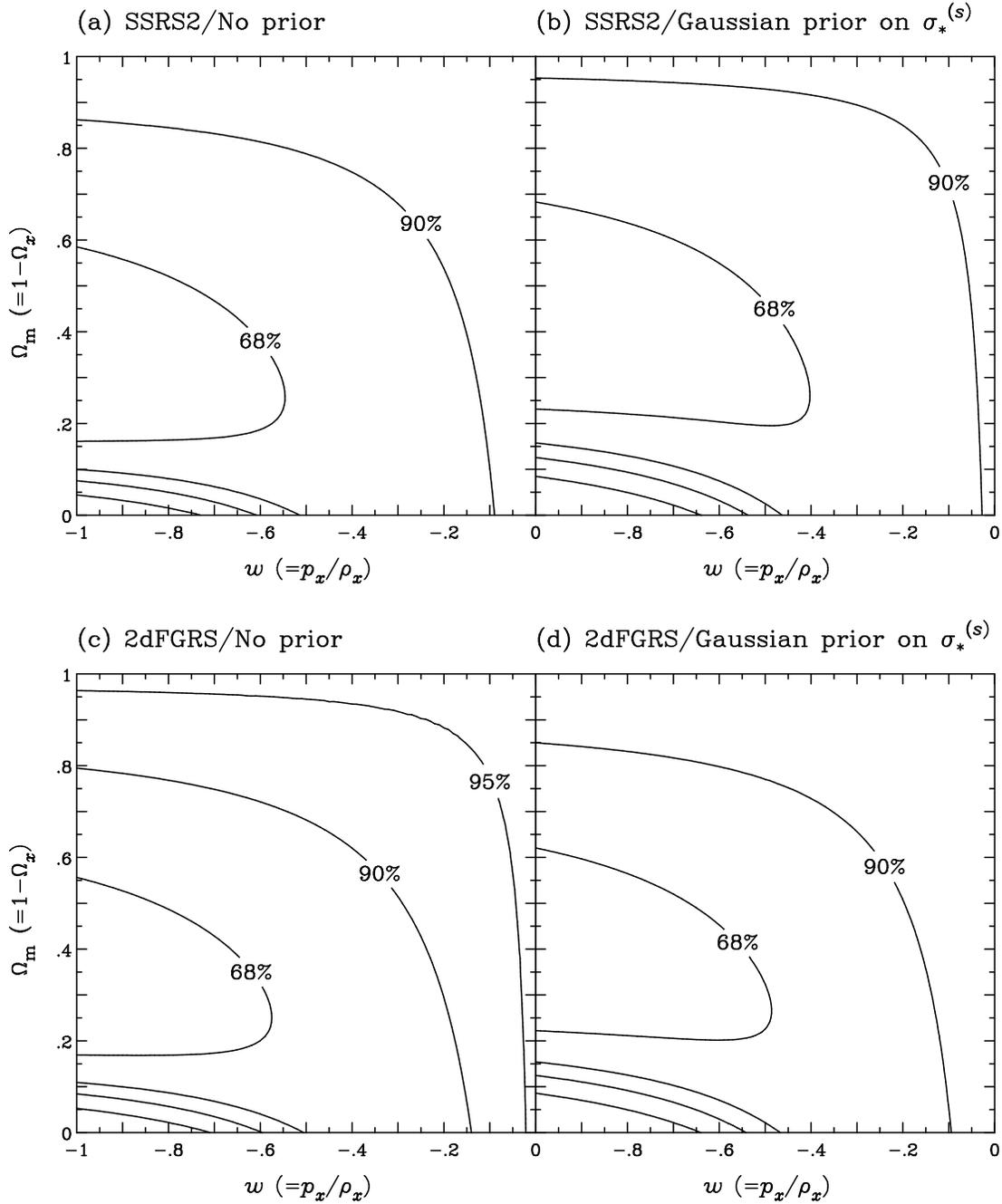}}
\end{picture}
\caption{
Likelihood regions in the $w$-$\Omega_{\rm m}$ plane with the prior that
the universe is flat for the same models illustrated in {Fig.}~6.
}
\label{}
\end{center}
\end{figure*}

{Fig.}~7 shows the $w$-$\Omega_{\rm m}$ plane in flat cosmology. 
As in {Fig.}~6, four different results are presented in {Fig.}~7 depending 
on the type-specific LFs and the assumption on the late-type characteristic 
velocity dispersion.
In Table~4, we summarize the 68\% and the 95\% confidence limits on the 
cosmological parameters.
\begin{table*}
\caption{Constraints on cosmological parameters for the four models 
illustrated in {Fig.}~6 and {Fig.}~7.}
\begin{tabular}{clllllllll} \hline
   & $\Omega_{\Lambda}-1.2\Omega_{\rm m}$  &   & 
  & $\Omega_{\rm m}(=1-\Omega_{\Lambda})$  &   &   
  & $w$  &   &   \\  
Model &  MLE$^1$ & 68\% limit & 95\% limit  &  MLE$^1$ & 68\% limit 
& 95\% limit  &  MLE$^1$  & 68\% limit & 
95\% limit \\ \hline
 (a) & 0.40  & $(-0.16,0.73)$ & $(-1.17,0.92)$  & 0.31  & (0.17,0.58) & 
  (0.08,1.06) & $-1$  &  $<-0.55$  &  ----- \\
 (b) & 0.12  & $(-0.50,0.50)$ & $(-1.52,0.74)$  & 0.40  & (0.24,0.68) & 
  (0.13,1.14) & $-1$  &  $<-0.41$  &  ----- \\
 (c) & 0.40  & $(-0.10,0.71)$ & $(-0.98,0.90)$  & 0.31  & (0.17,0.55) & 
  (0.09,0.96) & $-1$  &  $<-0.58$  &  $<-0.03$  \\
 (d) & 0.17  & $(-0.36,0.52)$ & $(-1.21,0.75)$  & 0.38  & (0.23,0.62)  &
  (0.13,1.00)  & $-1$  &  $<-0.49$  &  ----- \\
 \hline
\end{tabular}

$^1$ Maximum likelihood estimate which is the value minimizing the $\chi^2$
given by equation~(44).

\end{table*}
For the models considered in {Fig.}~6 and {Fig.}~7, the fitted values of 
the early-type characteristic velocity dispersion are somewhat different 
for the different early-type luminosity functions. Confidence limits on 
galactic parameters including the early-type characteristic velocity 
dispersion are presented next in section~4.2.

\subsection{Constraints on global properties of galaxy populations}
In deriving constraints on galactic parameters, cosmological 
parameters are varied to minimize the $\chi^2$ (equation~44) at each grid
point. Thus, the derived constraints on galactic parameters depend on the
allowed ranges of cosmological parameters. In this section, we only
consider flat cosmology (i.e.\ $\Omega_{\rm m}+ \Omega_x = 1$) with
$0 < \Omega_{\rm m} \le 1$ and fix the dark-energy equation of state $w$ at
$w = -1$ (which is the best-fit value when it is allowed to vary with the
constraint $w \ge -1$). 

{Fig.}~8 shows confidence limits in the parameter plane spanned by the 
early-type characteristic velocity dispersion $\sigma_*^{(e)}$ and the 
late-type characteristic velocity dispersion $\sigma_*^{(s)}$. The six 
different results in {Fig.}~8 are for the three different cases of the 
intrinsic shape distribution of galaxies, i.e.\ all oblate, one half oblate 
and the other half prolate, and all prolate, each for the SSRS2 and the 
2dFGRS type-specific LFs. 
\begin{figure*}
\begin{center}
\setlength{\unitlength}{1cm}
\begin{picture}(20,21)(0,0)
\put(0.,-1.){\includegraphics{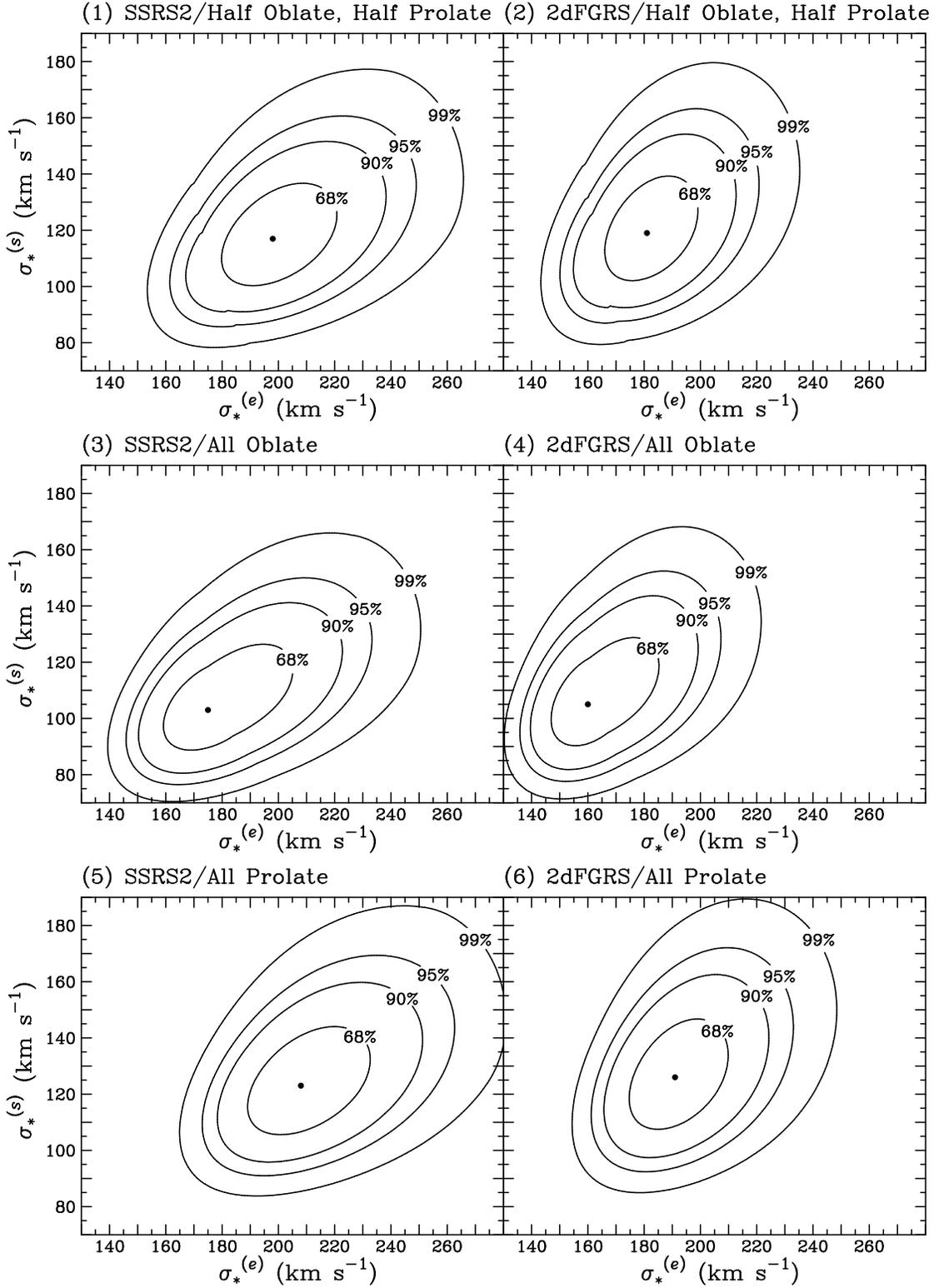}}
\end{picture}
\caption{
Likelihood regions in the $\sigma_{*}^{(e)}$-$\sigma_{*}^{(s)}$ plane
with the prior that the universe is flat, $w = -1$, and 
$0 < \Omega_{\rm m} \le 1$. Models (1), (3), and (5) are based on the 
SSRS2 type-specific LFs while models (2), (4), and (6) and based on the 
2dFGRS type-specific LFs. For models (1) and (2), one half of the galaxies
of each type are assumed to be oblate while the other half are assumed to be
prolate. For models (3) and (4) all galaxies are assumed to be oblate while 
for models (5) and (6) all galaxies are assumed to be prolate. 
}
\label{}
\end{center}
\end{figure*}
{Fig.}~9 shows the parameter plane spanned by the mean projected mass 
ellipticity of galaxies $\bar{\epsilon}$ and the early-type characteristic 
velocity dispersion $\sigma_*^{(e)}$ for the same six cases considered in 
{Fig.}~8.
\begin{figure*}
\begin{center}
\setlength{\unitlength}{1cm}
\begin{picture}(20,21)(0,0)
\put(0.,-0.8){\includegraphics{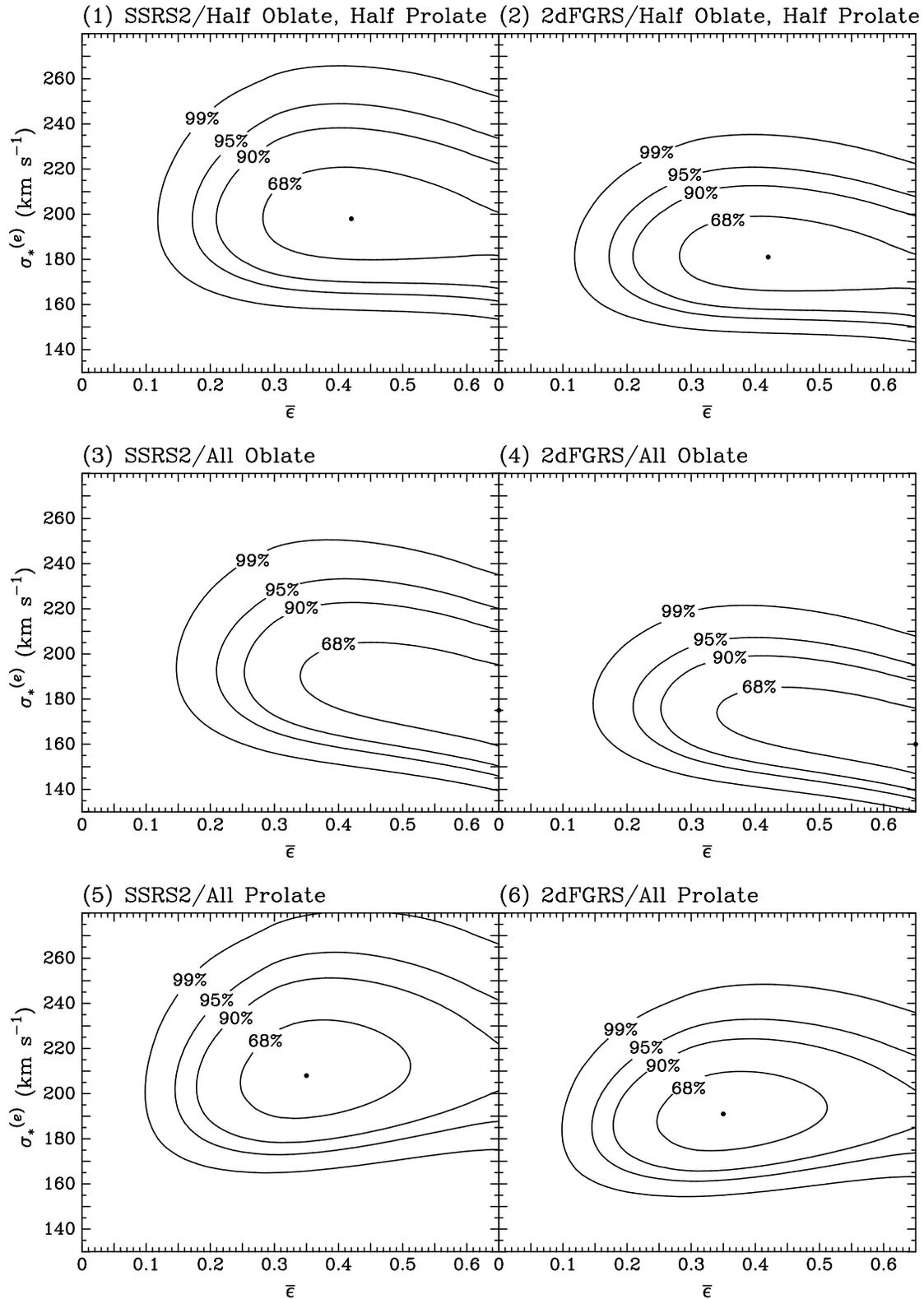}}
\end{picture}
\caption{
Likelihood regions in the $\bar{\epsilon}$-$\sigma_{*}^{(e)}$ plane for
the same models illustrated in {Fig.}~8.
}
\label{}
\end{center}
\end{figure*}
In Table~5, we summarize the 68\% and the 95\% confidence limits on the
galactic parameters.
\begin{table*}
\caption{Constraints on galactic parameters in flat cosmology for
the six models illustrated in {Fig.}~8 and {Fig.}~9.}
\begin{tabular}{clllllllll} \hline
   & $\sigma_{*}^{(e)}$  &    &   
  & $\sigma_{*}^{(s)}$  & 
 &   & $\bar{\epsilon}$  &  &   \\  
   & (km s$^{-1}$)  &    &   & (km s$^{-1}$)  &  &   &  &  &   \\  
 Model & MLE$^1$  &  68\% limit & 95\% limit & MLE$^1$  &  68\% limit 
& 95\% limit &  MLE$^1$  &  68\% limit & 95\% limit \\ \hline
 (1) & 198  & (180,220) & (162,248) & 117  & (101,136)  & (86,160) &
     0.42 & (0.28,---)  &  (0.17,---) \\
 (2) & 181  & (166,199) & (151,220) & 119  & (103,139)  & (87,163) &
     0.42 & (0.28,---)  &  (0.17,---) \\
 (3) & 175  & (160,205) & (146,233) & 103  & (89,126)  & (77,149) & 
     --- & (0.34,---)  &  (0.21,---) \\
 (4) & 160  & (147,185) & (136,207) & 105  & (91,128)  & (78,152) &
     --- & (0.34,---)  &  (0.21,---)  \\ 
 (5) & 208  & (190,232) & (173,262) & 123  & (106,144)  & (91,169) &
     0.35 & (0.25,0.51)  & (0.14,---)  \\
 (6) & 191  & (175,209) & (162,233) & 126  & (108,146)  & (93,172) & 
     0.35 & (0.25,0.51)  & (0.14,---)  \\ \hline
\end{tabular}

$^1$ Maximum likelihood estimate which is the value minimizing the $\chi^2$
given by equation~(44).
\end{table*}

\subsection{Systematic effects of varying the parameters for the source 
population}
The results presented in sections~4.1 and 4.2 were derived assuming the 
present best estimates of the redshift distribution and the differential 
number--flux-density relation for the CLASS sources as given in section~3.2. 
In other words, the present uncertainties in the redshift distribution and 
the differential number--flux-density relation for the CLASS sources were not
incorporated in the results in sections~4.1 and 4.2. Here we quantify 
the effects of varying the parameters used to describe the CLASS 
source population. The source parameters that can have significant 
impacts on the derived results on the cosmological
parameters are the mean redshift, $\bar{z}$ and the differential 
number--flux-density relation slope for flux densities lower than 30~mJy,
$\eta_2$ (section~3.2). Figure~10~(a) and (b) show respectively the effects
of varying the parameters  $\bar{z}$ and $\eta_2$ on the maximum likelihood
value of $\Omega_{\rm m}$ in flat cosmology with $w = -1$.
\begin{figure*}
\begin{center}
\setlength{\unitlength}{1cm}
\begin{picture}(13,12)(0,0)
\put(-2.5,13.5){\includegraphics{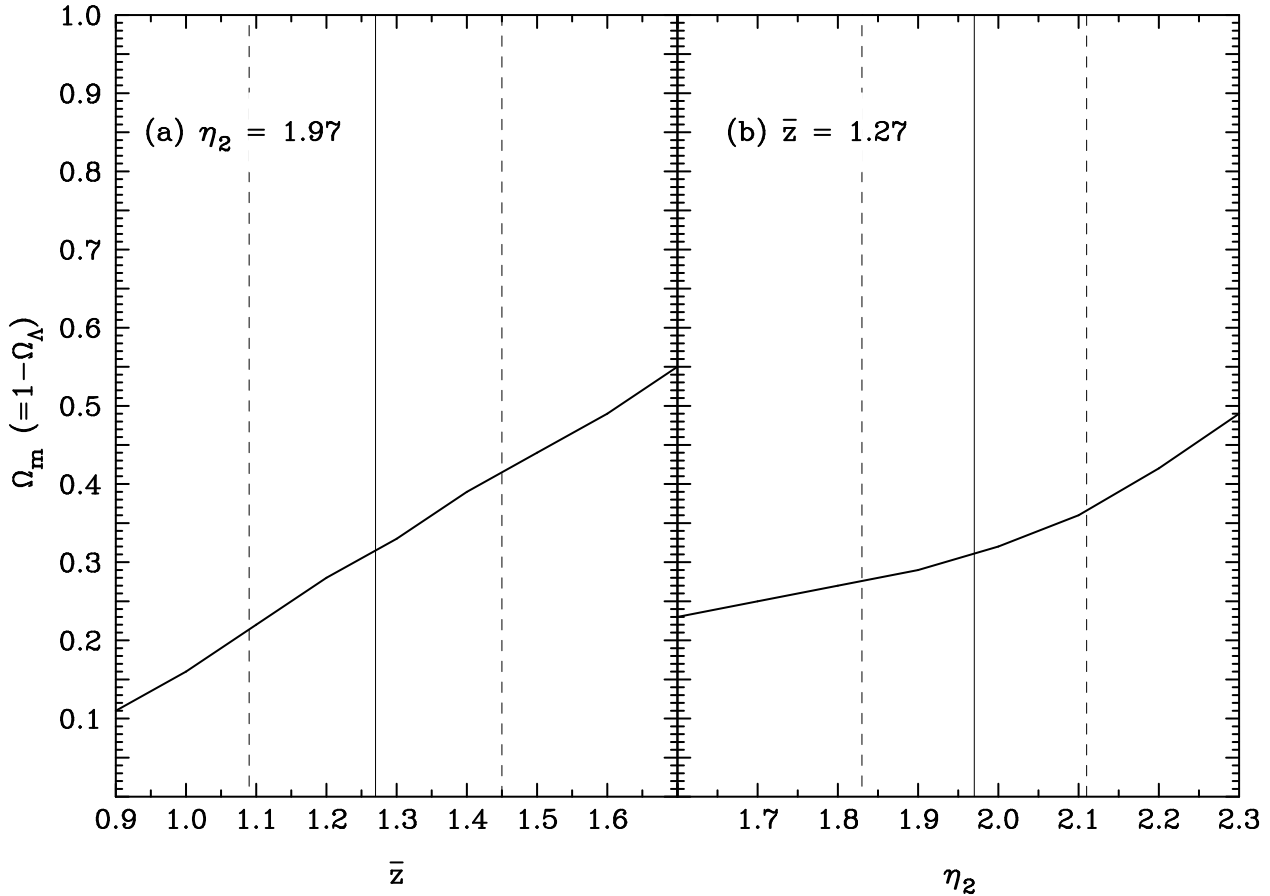}}
\end{picture}
\caption{
Systematic effects of varying (a) the mean source redshift,
$\bar{z}$ and (b) the $|dN/dS|$ slope for $S < 30$~mJy,
$\eta_2$, on the maximum likelihood value of $\Omega_{\rm m}$ in flat
cosmology with $w = -1$. The thin vertical solid lines, respectively, 
correspond to the present best estimates of $\bar{z}$ and $\eta_2$ while
the thin vertical dashed lines correspond to their present uncertainties.
}
\label{}
\end{center}
\end{figure*}

\section{DISCUSSION}

\subsection{Understanding the sensitivities of the model parameters
to the observed properties of statistical lensing}
The likelihood function [equation~41; or the $\chi^2$ function (equation~44)]
 for statistical lensing is a complicated function of
the parameters of the statistical lensing model (section~2.1.3). 
Moreover, the data are characterized by many properties
(section~3) which control the likelihood function. For these reasons, it is
not straightforward to see how the cosmological and galactic parameters are
constrained by the data. Here we discuss the sensitivities of the model
parameters to the properties of the data including numerical tests.

To help us see how the model parameters are constrained by the data, 
we rewrite the $\chi^2$ function (equation~44) as
\begin{equation}
\chi^2 = \chi^2_{\rm U} + \chi^2_{\rm L},
\end{equation}
where $\chi^2_{\rm U}$ and $\chi^2_{\rm L}$ are respectively the 
contributions from the unlensed sources and the lensed sources, which are 
given by
\begin{equation}
\chi^2_{\rm U}=
\sum_{k=1}^{N_{\rm U}}\ln\frac{1}{[1-\sum_m p_m^{\rm (all)}(k)]^2}
\end{equation}
 and
\begin{equation}
\chi^2_{\rm L}=
\sum_{l=1}^{N_{\rm L}}\ln\frac{1}{[\delta p_{m_l}^{\rm (one)}(l)]^2},
\end{equation}
where the integrated lensing probability $p_m^{\rm (all)}(k)$ and the
differential lensing probability $\delta p_{m_l}^{\rm (one)}(l)$ are given
by equations~(42) and (43). {From} equations~(51) and (52), we see that 
increasing (decreasing) lensing probabilities increases (decreases) 
$\chi^2_{\rm U}$ but decreases (increases) 
$\chi^2_{\rm L}$.\footnote{Quast \& Helbig (1999) used example
figures to demonstrate this point [see their Figs.\ (A1) and (A2)].}
The procedure of fitting the model parameters to the data is to minimize 
the sum of the oppositely behaving $\chi^2_{\rm U}$ and $\chi^2_{\rm L}$ by 
varying the lensing probabilities.

The lensing probabilities depend on many factors including galactic and
cosmological factors and the magnification bias factor which depends on the
source properties (section~2.1.2). In other words, the lensing probabilities 
are determined by the combination of all those factors. Accordingly,
the lensing probabilities can be varied in a number of different ways of
varying the factors. For example, the lensing probabilities can be decreased
by reducing the cosmological factors ($\approx \hat{D}(0,z)^2 |d\ell/dz|$), 
by decreasing $\sigma_{*}$, or even by reducing the cosmological factors
 and increasing $\sigma_{*}$
(in the last case, one of course has to reduce the cosmological factors
sufficiently enough to compensate for increasing $\sigma_{*}$). However, 
different combinations of the factors are distinguished by the 
differences in their capabilities of fitting the various properties of
the data. For example, combining a very large $\sigma_{*}$ and a very low 
value of the cosmological factors can give a `correct' lensing rate, but 
such a combination is discarded (i.e.\ it gives a relatively larger $\chi^2$
value) because it does not fit the observed image separation distribution. 

Roughly speaking, for a given dynamical normalisation (section~2.1.1),
the observed image separations constrain the characteristic velocity
dispersions $\sigma_{*}^{(e)}$ and $\sigma_{*}^{(s)}$ and the ratio
of the numbers of doubles and of quadruples constrains the surface density
ellipticity. For these constrained galactic factors and a magnification bias
factor determined from the estimated source properties, the cosmological 
factors are constrained mostly by the ratio 
of the numbers of the lensed sources and of the unlensed sources and (much)
less significantly by the redshifts of the lenses and of the sources.
In practice, the galactic and cosmological factors are, of course, 
simultaneously varied to minimize the total $\chi^2$. 

In Table~6, we present the calculated values of the  $\chi^2_{\rm U}$ 
(equation~51) and  $\chi^2_{\rm L}$ (equation~52) along with the fitted key 
parameters and the
predicted lensing rate for the four different models shown in {Fig.}~7 and 
Table~4 and two additional models to be compared. For models (a), (b), and 
(b$'$), the SSRS2 type-specific LFs are used while for models (c), (d), and
(d$'$), the 2dFGRS type-specific LFs are used. Notice that for models (b) and
(d) the late-type characteristic velocity dispersion is constrained using a
Gaussian prior, whose contribution to the total $\chi^2$ is, however, not
shown in Table~6. For the two additional models (b$'$) 
and (d$'$), the characteristic velocity dispersions are fixed at 
$\sigma_{*}^{(e)} = 225$~km~s$^{-1}$ and $\sigma_{*}^{(s)} = 145$~km~s$^{-1}$,
which are the preferred values in several previous analyses of statistical
lensing (e.g.\ Kochanek 1996a; Falco, Kochanek, \& Mu\~{n}oz 1998; Quast \&
Helbig 1999; Helbig et al.\ 1999).
\begin{table*}
\caption{Relative $\chi^2$ values and predicted lensing rates for various
models of statistical lensing. The $\chi^2$ is given by
$\chi^2=\chi^2_{\rm U}+\chi^2_{\rm L}$, where $\chi^2_{\rm U}$
(equation~51) and $\chi^2_{\rm L}$ (equation~52) are respectively the 
contributions from the unlensed sources and the lensed sources.
Model (a), (b), (c) and (d) are the four models illustrated in {Fig.}~7 and 
Table~4, here for the case of flat cosmology with a classical cosmological
constant. Notice that for the above models, the early-type and 
the late-type characteristic velocity dispersions were fitted. 
On the other hand, for Model~(b$'$) and (d$'$) the velocity dispersions
were fixed at $\sigma_*^{(e)} = 225$~km~s$^{-1}$ and 
$\sigma_*^{(s)} = 145$~km~s$^{-1}$ (see section~5.1). As for Model~(a) and 
(b), the SSRS2 type-specific LFs were used for Model~(b$'$) while as for
Model~(c) and (d), the 2dFGRS type-specific LFs were used for Model~(d$'$).
}
\begin{tabular}{cllllllll} \hline
 Model & $\Omega_{\rm m}$ & $\sigma_{*}^{(e)}$ & $\sigma_{*}^{(s)}$ & 
 $\chi^2_{\rm U}$ &  $\chi^2_{\rm L}$  
& $\chi^2=\chi^2_{\rm U}+\chi^2_{\rm L}$ & 
 $\Delta\chi^2$ & $\bar{p}^{a}$    \\  
   & ($=1-\Omega_{\Lambda}$) & (km s$^{-1}$) & (km s$^{-1}$) & 
  &  &   &   &   \\ \hline
 (a) & 0.31  & 197.8 & 116.7 & 25.17  & 186.96  & 212.12 & 1.53 &  1/711 \\
 (b) & 0.40  & 204.8 & 130.2 & 25.46  & 187.14  & 212.60 & 2.01 &  1/703 \\
 (b$'$) & 0.65 & 225.0 & 145.0 & 25.89  & 188.41  & 214.30 & 3.71 & 1/691 \\
 (c) & 0.31 & 180.8 & 119.0 & 25.08  & 185.52  & 210.59 & $\equiv 0$ &1/713\\
 (d) & 0.38 & 185.5 & 130.8 & 25.12  & 185.83  & 210.95 & 0.36 & 1/712\\
(d$'$) & 0.89 & 225.0 & 145.0 & 25.71 & 189.70 & 215.41 & 4.82& 1/696\\ \hline
\end{tabular}

$^{a}$ The observed lensing rate is 1/689.
\end{table*}
As can be seen in Table~6, galaxy populations following the 2dFGRS 
type-specific LFs with the fitted characteristic velocity dispersions
[Model~(c)] have the lowest value for the $\chi^2$. Compared with Model~(c),
Model~(a) in which galaxy populations follow the SSRS2 type-specific LFs
have a larger value for the $\chi^2$ by $\Delta \chi^2 = 1.53$. The larger
$\chi^2$ value for Model~(a) compared with Model~(c) is mostly due to the
increase in the value of $\chi^2_{\rm L}$. The implication for this is that 
Model~(a) does not fit the observed image separation distribution as well as
Model~(c) does. Likewise, other models presented in Table~6 have larger 
$\chi^2$ values mainly because of their larger values of the $\chi^2_{\rm L}$
compared with Model~(c). In particular, Models (b$'$) and (d$'$), in which the
characteristic velocity dispersions are held fixed, have significantly larger
$\chi^2_{\rm L}$ values compared with Model~(c) because the predicted mean
image separations are discrepant with the observed mean image separation for
those models. To show the above, we present in {Fig.}~11 
the predicted image separation distributions for 
Models (a), (b$'$), (c), and (d$'$). 
\begin{figure*}
\begin{center}
\setlength{\unitlength}{1cm}
\begin{picture}(14,12)(0,0)
\put(-3.5,14.5){\includegraphics{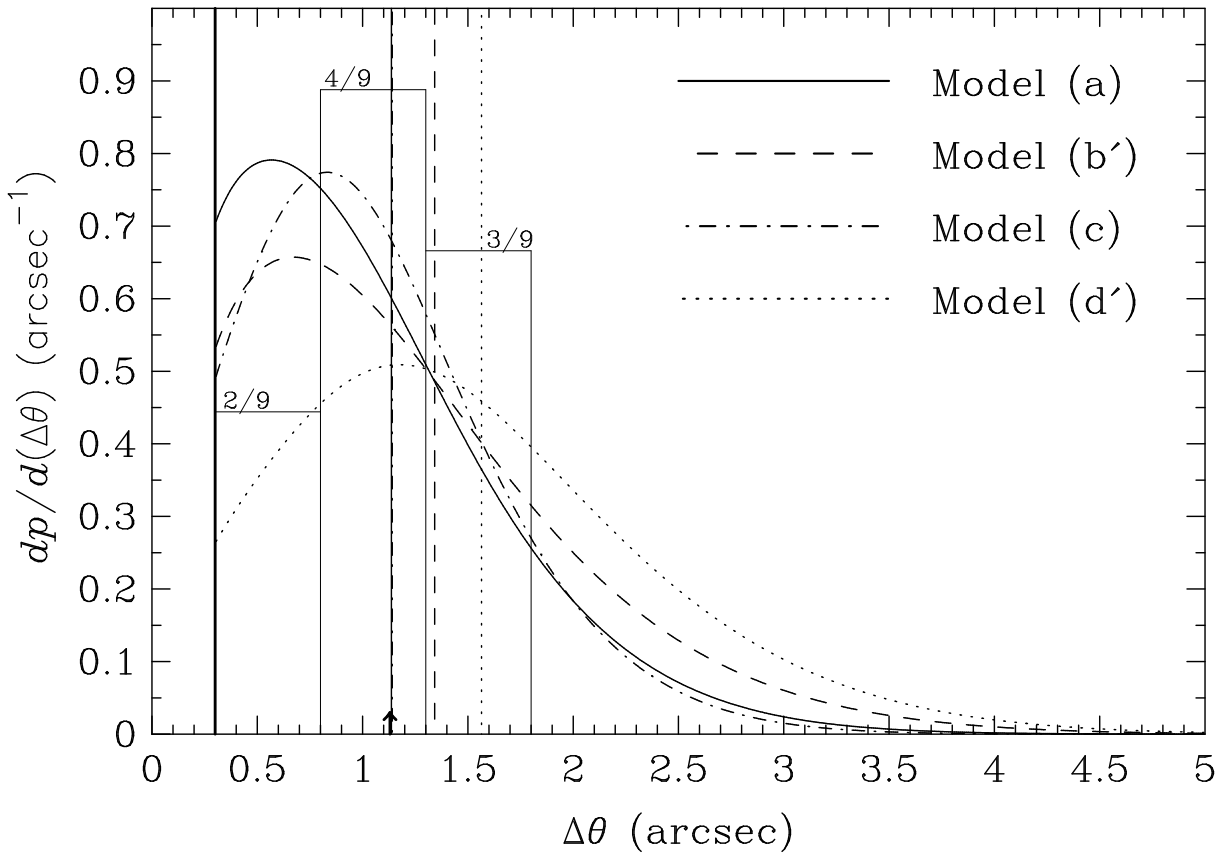}}
\end{picture}
\caption{
Image separation probability distributions for $\Delta\theta \ge 0.3$~arcsec
of multiply-imaged CLASS sources at redshift $z_s = 2$ with lens 
redshifts $0.3 \le z_l \le 1$ (assuming that the unmeasured lens redshifts
are also in this range). Theoretical distributions predicted 
by four models which can be found in Table~6 are 
overplotted on the observed distribution (i.e.\ the histogram) 
of the 9 multiply-imaged sources in the final CLASS statistical
sample whose image splittings are known (or likely) to be caused by single
galaxies. The small arrow marks the observed mean image separation, namely
$\langle \Delta\theta \rangle_{\rm CLASS} = 1.13$~arcsec, while the four
vertical lines denote the predicted mean image separations of the four models.
Notice that the mean image separations predicted by Model~(a) and (c), in
which the early-type and the late-type characteristic velocity dispersions
were fitted, agree well with the observed mean image separation 
(see section~5.1). The mean image separations predicted by Model~(b$'$) and 
(d$'$), in which the early-type and the late-type characteristic velocity 
dispersions were fixed at $\sigma_*^{(e)} = 225$~km~s$^{-1}$ and 
$\sigma_*^{(s)} = 145$~km~s$^{-1}$ (see section~5.1), are discrepant with
the observed mean image separation. The distribution predicted by Model~(c),
in which the 2dFGRS type-specific LFs were used, agrees best with the
histogram (this explains why the model has the lowest $\chi^2$ value), whose
bins, however, consist only of small numbers of data points.
}
\label{}
\end{center}
\end{figure*}
In {Fig.}~11 are also displayed the model predicted mean image 
separations and the observed mean image separation for the multiply-imaged
systems whose image separations are used in the fit (see section~3.1). 
Finally, notice that all the models in Table~6 predict correct lensing rates
despite the differences in their capabilities of fitting the observed image 
separation distribution. This is, of course, the consequence of adjusting
the value of $\Omega_{\rm m}$ to fit the observed lensing rate for the fitted
(or given) galactic parameters in each model.

{From} the above examination of the example models, we can summarize how the
galactic and cosmological parameters are constrained by the data as follows.
For the given type-specific LFs, the observed image separations constrain
the characteristic velocity dispersions through the $\chi^2_{\rm L}$ term
and then the observed lensing rate constrains cosmological parameters
through the sum of the $\chi^2_{\rm L}$ and the $\chi^2_{\rm U}$ terms. Since
the lensing rate can be fitted by adjusting cosmological parameters for any
given galactic parameters, the lensing rate alone cannot constrain 
cosmological parameters unless one has {\it a priori} accurate knowledge of
the galactic parameters. 

\subsection{Comparison with previous results in statistical lensing}
Previous analyses of statistical lensing were based on an 
optically-selected sample, partial samples from radio-selected
CLASS sources and other radio-selected sources, and galaxy luminosity 
functions derived prior to the 2dFGRS and the SDSS observations.
The optically-selected sample used for previous analyses  
is summarized in Kochanek (1996a) and consists of a combined sample 
of 862 highly luminous QSOs, out of which 5 QSOs are multiply-imaged due to 
galactic mass-scale lenses. Radio-selected multiply-imaged sources used for 
previous analyses include 4 multiply-imaged flat-spectrum radio sources 
(e.g.\ Helbig et al.\ 1999) out of a total of 6 JVAS multiply-imaged sources 
(Browne et al.\ 2003) and several other multiply-imaged sources including 
CLASS sources. 

Based on the optically-selected sample by Kochanek (1996a) and radio-selected 
samples, some previous analyses of statistical lensing (Kochanek 1996a; 
Falco, Kochanek, \& Mu\~{n}oz 1998; Quast \& Helbig 1999)
obtained limits on $\Omega_{\Lambda}$ in flat cosmology (or, on 
$\Omega_{\Lambda} - \Omega_{\rm m}$) that would rule out a high 
 $\Omega_{\Lambda}$. Most notably, Kochanek (1996a)
obtained $\Omega_{\Lambda} \la 0.66$ a 95\% confidence with a maximum
likelihood value of $\Omega_{\Lambda} \sim 0$. This result is significantly 
different from the results on cosmological parameters that are derived in 
section~4.1. To understand the results by Kochanek (1996a)
and the others mentioned above, it is important to notice that 
(1) they used for the early-type characteristic velocity dispersion
$\sigma_{*}^{(e)} = 225 \pm 22.5$ km~s$^{-1}$ from Kochanek (1994) as a
prior information and (2) this prior value of $\sigma_{*}^{(e)}$ was also 
consistent with the observed image separations of the multiply-imaged sources
used by them for their adopted early-type luminosity function 
(Kochanek~1996a). The latter point is the consequence of combining two 
things. First, the mean image separation 
for the 5 optically-selected lens systems is 
$\langle\Delta\theta\rangle_{\rm optical} = 1.60$ arcsec which is 
significantly larger than the mean separation of 
$\langle\Delta\theta\rangle_{\rm CLASS} = 1.13$ arcsec for the 9 
multiply-imaged sources (or, 
$\langle\Delta\theta\rangle_{\rm CLASS}^{(e)} = 1.23$
arcsec excluding the confirmed spiral lens system 0218+357)
in the final CLASS statistical sample (section~3.1) whose image separations
are (likely to be) due to single galactic potentials.\footnote{
The radio-selected multiply-imaged sources collected by Kochanek (1996a) and 
used to constrain $\sigma_{*}^{(e)}$ have a mean separation of 
$\langle\Delta\theta\rangle_{\rm radio(Kochanek)} \approx 1.6$ which is 
nearly equal to $\langle\Delta\theta\rangle_{\rm optical}$.} Second, 
the fitted value of $\sigma_{*}^{(e)}$ is a function of the early-type
faint-end slope $\alpha^{(e)}$ (for a fixed Faber-Jackson exponent 
$\gamma_{\rm FJ}$) as seen in section~4.2, and the early-type LF used by
the above authors has a relatively steep faint-end slope of 
$\alpha^{(e)}= -1.00\pm 0.15$ similar to that for the SSRS2 early-type LF 
(section~3.3) and thus requires a relatively larger value of 
$\sigma_{*}^{(e)}$ (compared with a shallower faint-end slope as in the 
2dFGRS early-type LF). In passing, we mention that 
the local early-type characteristic number density $n_{*,0}^{(e)}$
estimated by Kochanek (1996a) agrees well with that
for the SSRS2 early-type LF.

Like the above authors, Helbig et al.\ (1999) adopted 
$\sigma_{*}^{(e)} = 225 \pm 22.5$ km~s$^{-1}$ in deriving limits on 
cosmological parameters based on their JVAS sample. Helbig et al.\ (1999)
obtained $-0.85 < \Omega_{\Lambda} < 0.84$ at 95\% confidence with a maximum
likelihood value of $\Omega_{\Lambda} > 0$ for a flat cosmology. This 
result is somewhat different from that by Kochanek (1996a) in the sense that
it marginally favors a positive  $\Omega_{\Lambda}$. However, it is also
different from the results of this paper (section~4.1) which favor a 
$\Omega_{\Lambda}$-dominated universe. Given the difference between the value
of $\sigma_{*}^{(e)}$ adopted by Helbig et al.\ (1999) and the values 
fitted to CLASS lenses (section~4.2), it is perhaps not surprising to find
that the Helbig et al.\ (1999) result and our results are different. 
Yet, it would be of interest to assess and understand the difference
between the Helbig et al.\ (1999) result and our results more concretely
and clearly, especially since the JVAS sample used by Helbig et al.\ (1999)
is a CLASS subsample just as the final CLASS statistical sample is. 
What if $\sigma_{*}^{(e)}$ was fitted to the 
lenses in the Helbig et al.\ (1999) sample rather than being fixed at 
$\sigma_{*}^{(e)} = 225$ km~s$^{-1}$?  
What factors other than the value of $\sigma_{*}^{(e)}$
(in particular, the more realistic statistical lensing model, 
the adopted profile likelihood ratio method of estimating parameters
and their errors, and sample variance) might be attributed to the difference
between the results of Helbig et al.\ (1999) and of this paper?
First of all, we recall that there
is no significant difference in galactic parameters other than 
$\sigma_{*}^{(e)}$ between Helbig et al.\ (1999) and this paper. 
To help us see separately the effects of varying statistical lensing model, 
parameter and error estimation method, and sample, 
we consider {Fig.}~12 which shows four variants of {Fig.}~6~(a) 
(i.e., results based on the SSRS2 type-specific LFs).
{Fig.}~12~(a) and (b) are the results based on the same data used by Helbig
et al.\ (1999)\footnote{A noticeable difference in sample selection criteria
between the Helbig et al.\ (1999) sample and the final CLASS statistical
sample is that the lower limit of the fainter-to-brighter image flux-density
ratio for doubly-imaged sources is 0.05 for the Helbig et al.\ (1999) sample.
This was taken into account in deriving the results shown in {Fig.}~12~(a)
and (b).} but using our model. {Fig.}~12~(a) is the result obtained 
considering only early-type galaxies and taking 
$\sigma_{*}^{(e)}=225$ km~s$^{-1}$ [i.e., adopting the same approach used
by Helbig et al.\ (1999)]. To obtain {Fig.}~12~(a), we also assumed that 
oblates and prolates have equal frequencies (i.e.\ dynamical normalisation
$\lambda \sim 1$) and took a mean projected galaxy mass ellipticity
$\bar{\epsilon} = 0.42$ which gives the best-fit to the final CLASS 
statistical sample (see {Fig.}~9). {Fig.}~12~(b) is the result obtained by 
applying our new approach adopted in this paper to the Helbig et al.\ (1999)
data. This means that both early-type and late-type populations were
considered and their characteristic velocity dispersions were fitted to the
observed image separations of the Helbig et al.\ (1999) 4 multiply-imaged
systems. {Fig.}~12~(c) and (d) are the results based on two subsamples of the
final CLASS statistical sample (section~3.1) and using the same model and
approach adopted in this paper. {Fig.}~12~(c) is based only on the JVAS
subsample within the final CLASS statistical sample which comprises 1749
sources including three multiply-imaged sources 0218+357, 1422+231, and
2114+022. {Fig.}~12~(d) is based on a subsample of 2308 sources
including four multiply-imaged sources
0445+123, 0712+472, 2045+265, and 2319+051 which were randomly selected from
the subsample of 7209 sources including 10 multiply-images sources
within the final CLASS statistical sample that are not JVAS sources.
\begin{figure*}
\begin{center}
\setlength{\unitlength}{1cm}
\begin{picture}(18,18)(0,0)
\put(0.,-2.){\includegraphics{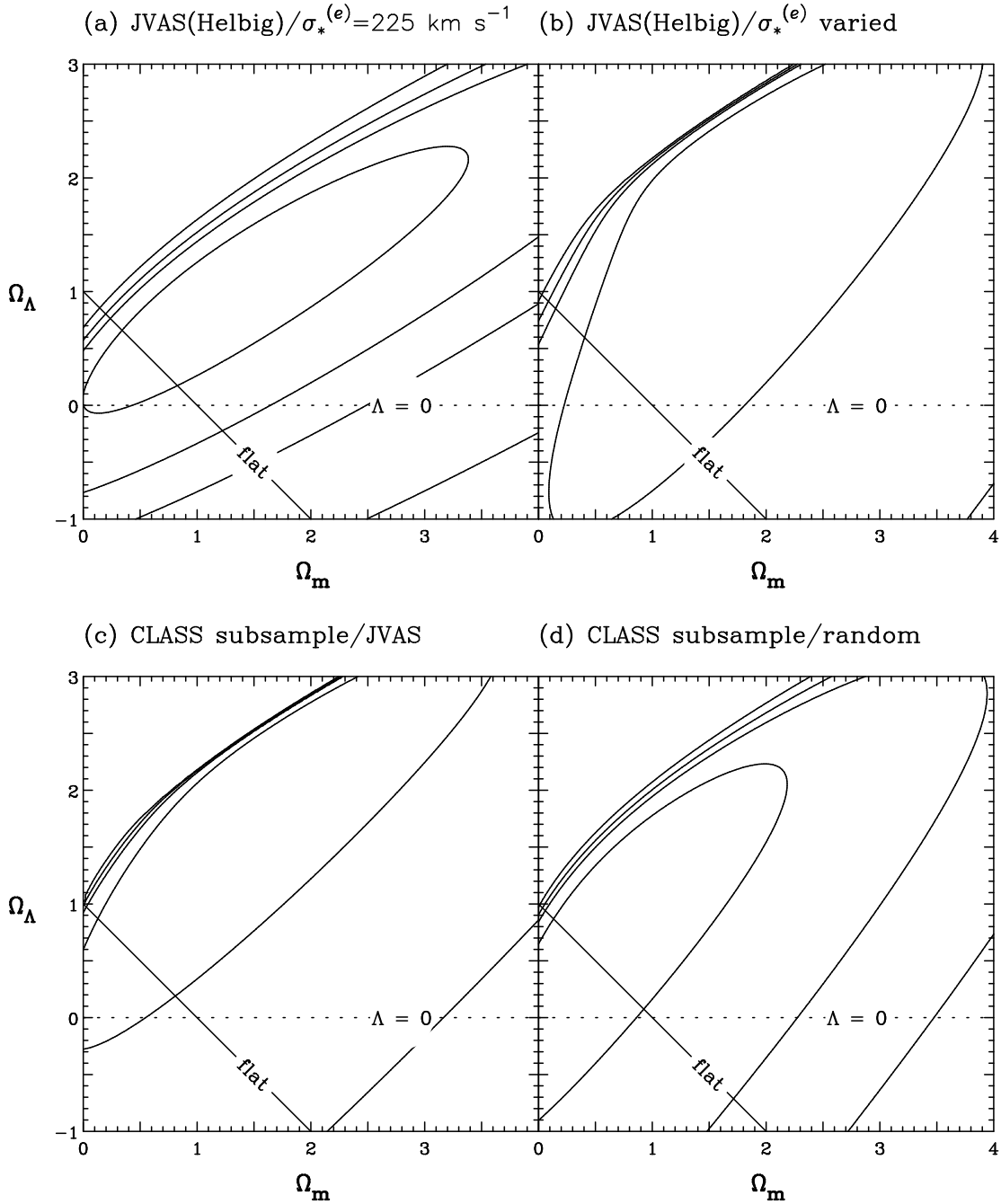}}
\end{picture}
\caption{
Likelihood regions in the $\Omega_{\rm m}$-$\Omega_{\Lambda}$ plane for the
SSRS2 type-specific LFs as in {Fig.}~6(a) based on previous and present
subsamples from the entire JVAS and CLASS sources, shown to illustrate the
effects of different statistical lensing models, different parameter and
error estimate methods, and different samples. The four contours in each
panel, respectively, correspond to the 68\%, 90\%, 95\%, and 99\% confidence
limits for one parameter. (a) \& (b) are based on the Helbig et al.\ (1999)
sample; (a) is the result of only fitting the lensing rate fixing early-type
characteristic velocity at $\sigma_*^{(e)} = 225$~km~s$^{-1}$ whereas (b) is
the result of simultaneously fitting both the lensing rate and the image
separations. Both (c) and (d) are the results of simultaneously fitting
both the lensing rate and the image separations based on subsamples of
the final CLASS statistical sample: (c) is 
the result based on the subsample 
that comprises only 1749 JVAS sources whereas (d) is based on a random 
subsample that comprises 2308 non-JVAS sources. 
The comparison of (a) and {Fig.}~1 of Helbig et al.\ (1999)
shows the effects of the different statistical lensing models.
The comparison of (a) and (b) shows the effects of the different parameter
and error estimate methods. The comparison among (b), (c), and (d) shows 
differences or similarities among samples. 
Finally, the comparison of (d) [or (c)] and {Fig.}~6(a) shows
the effects of different sample sizes.
See section~5.2 for a full discussion of the results.
}
\label{}
\end{center}
\end{figure*}

Let us briefly discuss the implications of the results displayed in
{Fig.}~12. {From} {Fig.}~12~(a), we have
$\Omega_{\Lambda} = 0.48^{+0.16}_{-0.30}$ (68\%) 
and $0.48^{+0.32}_{-0.98}$ (95\%) for a flat cosmology. 
This result favors a positive $\Omega_{\Lambda}$ universe
and the difference from the Helbig et al.\ (1999) result is not significant.
Thus, we are led to conclude that the differences in the details of the
statistical lensing model (in particular, cross sections and magnification
biases) between this paper and Helbig et al.\ (1999) make little difference
in deriving limits on cosmological parameters based on presently available
data (i.e.\ within presently reachable accuracies). 
{From} {Fig.}~12~(b), we have 
$\Omega_{\Lambda} = 0.32^{+0.26}_{-0.72}$ (68\%) for a flat cosmology. 
The somewhat lower maximum-likelihood estimate of $\Omega_{\Lambda}$ from 
{Fig.}~12~(b) compared with that from {Fig.}~12~(a) is the consequence of 
the early-type characteristic velocity dispersion larger than 225~km~s$^{-1}$ 
fitted to the image separations that are due to early-type galaxies
in the Helbig et al.\ (1999) sample (the mean image separation for 0414+054,
1030+074, and 1422+231 is 1.64 arcsec). Another main difference between
{Fig.}~12~(a) and (b) is that the confidence interval at a given confidence
level is much smaller in {Fig.}~12~(a) although both were based on the same
data. This is, of course, due to the fact that $\sigma_*^{(e)}$ was fixed for
{Fig.}~12~(a) while it was varied for {Fig.}~12~(b) based on the profile 
likelihood ratio method (see section 4.1). The above comparison of 
{Fig.}~12~(a) and (b) shows that different approaches of treating galactic
parameters can make significant differences in the derived values
of cosmological parameters for the same data. The results displayed in
{Fig.}~12~(c) and (d) are based on the profile likelihood ratio method as in
{Fig.}~12~(b). We have $\Omega_{\Lambda} = 0.68^{+0.18}_{-0.48}$ and 
$0.64^{+0.22}_{-0.56}$ at 68\% confidence respectively from {Fig.}~12~(c) 
and (d). These values of $\Omega_{\Lambda}$ are very similar to
each other but different from the value obtained from {Fig.}~12~(b).
For example, the confidence interval of $\Omega_{\Lambda}$ implied by the
Helbig et al.\ (1999) data excludes at 68\% confidence both of the
maximum-likelihood estimates of $\Omega_{\Lambda}$ shown on {Fig.}~12~(c) and
(d). In other words, even if not at a high statistical significance,
the Helbig et al.\ (1999) data are inconsistent with two subsets of the
the final CLASS statistical sample when all three of the data sets are
analysed and interpreted in the same manner. 
This could imply that the Helbig et al.\ (1999) sample is incomplete
within their specified sample definitions or is not as observationally
well-defined as the final CLASS statistical sample. Alternatively, 
the Helbig et al.\ (1999) sample simply (as a rare occurrence) 
happens to include the lensed system 0414+054 that has an image
separation of 2.09~arcsec and is a main factor behind the low value of
$\Omega_{\Lambda}$.
Comparing {Fig.}~12~(d) [or {Fig.}~12~(c)] with {Fig.}~6~(a), we find that as
the sample size becomes four times larger, the 68\% confidence interval
approximately halves consistent with the expectation for a Poisson error. 

In summary, Kochanek (1996a), Helbig et al.\ (1999) and the other authors
above either found or adopted $\sigma_{*}^{(e)} \approx 225$ km~s$^{-1}$,
and for this and other adopted galactic parameters the observed lensing rates
in the optical sample of QSOs and the JVAS samples of flat-spectrum radio
sources were most consistent with relatively low values of
$\Omega_{\Lambda}$. In particular, some of these authors produced results
that even appeared to rule out a high value of $\Omega_{\Lambda}$ such as
$\Omega_{\Lambda} = 0.7$.

Some previous analyses of statistical lensing were based on choices of
galactic parameters that were different from those used by Kochanek (1996a),
Helbig et al.\ (1999), and the others above. 
Cooray (1999) used choices of $n_{*,0}^{(e)}$ and 
$\alpha^{(e)}$ that were similar to those used by Kochanek (1996a) but 
separated the early-type population into an elliptical (E) population and an
S0 population with a number ratio of E: S0 = 1: 1.6 and estimated the 
characteristic velocity dispersions from his chosen LFs and Faber-Jackson 
relation as $\sigma_{*} = 210_{-11}^{+10}$ km~s$^{-1}$ for the E population
and $\sigma_{*} = 194_{-10}^{+12}$ km~s$^{-1}$ for the S0 population.
The Cooray (1999) estimates of the E and S0 characteristic velocity 
dispersions correspond to an effective characteristic velocity dispersion for
the entire early-type population that is $\ga 1 \sigma$ lower than the
estimate by Kochanek (1994). Not surprisingly, Cooray (1999) obtained a
relatively high upper limit of $\Omega_{\Lambda} \la 0.79$ 
at 95\% confidence in flat cosmology from the lensing rate in an 
incomplete subsample of $\sim 6500$ CLASS sources. Notice, however, that 
Cooray (1999) did not use the observed image separations in his analysis
to test whether the chosen LFs and characteristic velocity dispersions 
were consistent with them or not. 

Choices of galactic parameters that are significantly 
different from those used by Kochanek (1996a), Cooray (1999), 
and the others above were considered by Chiba \& Yoshii (1999) and
Cheng \& Krauss (2001). In particular, based on the same lens systems used
by Kochanek (1996a), Chiba \& Yoshii (1999) found $\Omega_{\Lambda} \ga 0$
at 98\% confidence with a maximum likelihood value of 
$\Omega_{\Lambda} \approx 0.7$ because of their radically different
choices of galactic parameters. The galactic parameters preferred by 
Chiba \& Yoshii (1999) and Cheng \& Krauss (2001) can be characterized by
a very shallow or inverted (i.e.\ positive) faint-end slope for the 
early-type LF. Furthermore, Cheng \& Krauss (2001) argued for a  shallow
slope of $\gamma \approx 2$ to 3 in the $\sigma_{\rm DM}$-$L$ relation for
the early-type galaxies, and Chiba \& Yoshii (1999) used an early-type 
characteristic number density [i.e.\ an early-type LF by Loveday et al.\ 
(1992)] that is a factor of $\approx 2$ to 3 lower than those in the SSRS2
and the 2dFGRS early-type LFs (Table~3). The shallow/inverted 
early-type faint-end slopes necessarily require lower values for 
$\sigma_{*}^{(e)}$ for the given same image separations compared with
a steeper slope (as noted above). The high value for $\Omega_{\Lambda}$
obtained by Chiba \& Yoshii (1999) was the combined effect of the 
significantly lower values for the early-type characteristic velocity
dispersion and number density than those used by Kochanek (1996a). 
Although the limits on $\Omega_{\Lambda}$ obtained by Chiba \& Yoshii (1999)
agree well with the results on cosmological parameters based on the final 
CLASS statistical sample (section~4.1), we note that 
(1) the early-type LF used by Chiba \& Yoshii (1999) is discrepant with
present more reliable early-type LFs (section~3.3) and (2) the statistical
lensing samples used by Chiba \& Yoshii (1999) (i.e.\ those collected by 
Kochanek 1996a) were defined using selection functions that
are observationally not as reliable as those used to
define the final CLASS statistical sample and thus are more 
likely to be subject to unknown biases (which may explain 
the difference in the mean image separations between the final CLASS 
statistical sample and other samples). 

Compared with the observational information that was available to the 
previous analyses of statistical lensing (in particular those that put
limits on cosmological parameters), the presently available observational 
information used in this work is more abundant and more reliable. 
The final CLASS statistical sample is defined so as to satisfy the 
selection criteria (section~3.1) that are both observationally well-defined 
and reliable. This means that its biases are well understood and can be 
accurately taken into account in the analysis of the sample. [For comparison,
optically-selected samples suffer from dust extinction which cannot be
accurately taken into account for analyses of statistical lensing 
(see Falco et al.\ 1999).] The final CLASS statistical sample is the 
most significant piece of information that has been available only to this 
work. The total galaxy LFs derived from the 2dFGRS and the SDSS observations
and more recently derived type-specific LFs have now significantly reduced 
uncertainties in the early-type LF. The fact that the well-defined final CLASS
statistical sample and the relatively more reliable early-type LFs were used 
and all the factors in the statistical lensing model were carefully taken into
account based on relatively reliable data and with thorough understanding, 
makes the results on cosmological parameters reported in Chae et al.\ (2002)
and here (section~4.1) much more reliable than those from previous analyses.

So far in this section, we have focused on cosmological parameters in the 
comparison with previous analyses of statistical lensing. 
We now compare our derived results on galactic parameters with those 
in previous analyses. As emphasized above, the mean image separation in the
final CLASS statistical sample is significantly smaller than those in the
optical QSO sample and the heterogeneously defined radio-selected samples
(Kochanek 1996a). We recall that this was the main reason for the difference
in the cosmological constraints between Kochanek (1996a) and this work 
(section~4.1). The implication for the early-type characteristic velocity 
dispersion is that it is $\sim 27$  km~s$^{-1}$ lower than the estimate
by Kochanek (1994) for the steep SSRS2 early-type LF and 
$\sim 44$ km~s$^{-1}$ lower for the shallow 2dFGRS LF.\footnote{This is 
based on the dynamical normalisation for the equal numbers of 
the oblates and the prolates, i.e.\ $\lambda(f) \approx 1$.} We point out that
this fitted value of the early-type characteristic velocity dispersion is
also consistent with our estimate of the early-type characteristic magnitude
(section~3.3). Namely, our estimate of the early-type characteristic 
magnitude $M_{*,0}^{(e)}(B)$ is $\sim 0.5$~mag fainter than the one 
adopted by Kochanek (1996a) and it implies $\sigma_{*,0}^{(e)} \sim 190$
km~s$^{-1}$ either by a Faber-Jackson relation (de~Vaucouleurs \& Olson 1982)
or a $\sigma_{\rm DM}$-$L$ relation (Kochanek 1994) (see section~4).

Our statistical lensing model incorporates the (differential) 
lensing probabilities of specific image multiplicities for the multiply-imaged
sources using the SIE lens model (rather than a circular lens).
This allows us to constrain the mean projected mass ellipticity of galaxies
from the multiply-imaged sources of well-defined image multiplicities.
In section~3.1, we pointed out that 10 systems out of the 13 lens systems in
the CLASS statistical sample have well-defined image multiplicities for which
single galactic potentials are (likely to be) only responsible. 
The 10 systems contain 6 doubly-imaged ones 
and 4 quadruply-imaged ones. The relative frequency of
the doubles in the final CLASS statistical sample is similar to 
(or slightly higher than) those in the samples used by King \& Browne 
(1996), Kochanek (1996b), and Keeton, Kochanek, \& Seljak (1997). However,
the relative frequency for the doubles is higher than that in the CLASS 
subsample used by Rusin \& Tegmark (2001), in which there are 5 doubles and 7 
quadruples.\footnote{Rusin \& Tegmark (2001), as in this work, did not use
compound lenses in which multiple potentials are responsible for the image
multiplicities.}  This difference is for two reasons. First, in the time 
since Rusin \& Tegmark (2001) did their analysis based on CLASS lens systems 
available at that time, the CLASS group has identified two new lens systems
0445+123 (Argo et al.\ 2003) and 0631+519, both of which are doubly-imaged
systems and are now included in the final CLASS statistical sample. 
Second, the observational criteria used to define the final CLASS statistical 
sample (section~3.1) are not only more stringent but more reliable
than those used to define the statistical
sample used by Rusin \& Tegmark (2001). Based on their statistical sample
and the SIE lens model, Rusin \& Tegmark (2001) concluded that the predicted 
mean mass ellipticity of the early-type galaxies 
($\bar{\epsilon}_{\rm mass} \approx 0.6$)
was much higher than an observed mean light ellipticity . 
In particular, Rusin \& Tegmark (2001) found that the observed ellipticity 
distribution of the early-type galaxies in the Coma cluster 
was inconsistent with the relative frequencies of the doubles and the 
quadruples in their sample at 98\% statistical significance. 

Based on the final CLASS statistical sample and the SIE lens model,
we find that the best-fit mean mass ellipticity for the (early-type) 
galaxies\footnote{Recall that a common mean mass ellipticity for both
early-type and late-type populations of galaxies is considered and fitted to
the image multiplicities that are generated by both early-type and late-type 
(appropriately isolated) lensing galaxies. Since most of the lensing galaxies
are early-type, the common mean mass ellipticity can be regarded as the mean
mass ellipticity of early-type galaxies with little error.}
is $\bar{\epsilon}_{\rm mass} = 0.42$ ({Fig.}~9; Table~5) and the 68\% lower
limit is $0.28$ for the dynamical normalisation of the equal frequencies
of the oblates and the prolates. For the dynamical normalisation of all 
oblates, the 68\% lower limit is increased to 0.34. For the dynamical 
normalisation of all prolates, 
$\bar{\epsilon}_{\rm mass} = 0.35_{-0.10}^{+0.16}$ at 68\%.
These results are consistent with the observed mean light 
ellipticity $\bar{\epsilon}_{\rm light,DEEP/GSS} \approx 0.38$ ({Fig.}~18 in
Im et al.\ 2002) for the field early-type galaxies of the Deep Extragalactic
Evolutionary Probe (DEEP) Gross Strip Survey (GSS) or 
$\bar{\epsilon}_{\rm light,Coma} \approx 0.32$ for the combined E and S0
galaxies in the Coma cluster [{Fig.}~3(a) in J$\o$rgensen \& Franx 1994]. 
Therefore, the potential `ellipticity crisis' for gravitational lenses that
were noticed and investigated in the previous works (Rusin \& Tegmark 2001;
Kochanek 1996a; King \& Browne 1996) based on samples that are not as 
well-defined as the final CLASS statistical sample, has now significantly
weakened. 

\subsection{Possible sources of systematic errors}

\subsubsection{Mass profiles, intrinsic shapes, and velocity
dispersions of galaxies}
In this work, we adopted the singular isothermal mass profile and
the axisymmetric intrinsic shapes (i.e., oblate and prolate) to model 
galaxy lenses. We also assumed that the distribution function of a galaxy 
depends only on the relative energy and the angular momentum component 
parallel to the symmetry axis of the galaxy. With the adopted mass model 
and the assumption on the distribution function, 
we calculated, as functions of apparent axial ratio and 
line-of-sight velocity dispersion, the multiple-imaging cross sections
averaged over all the combinations of intrinsic shapes and inclination angles
that are consistent with the apparent axial ratio for the assumed relative 
frequencies of oblates and prolates. The apparent axial ratio and the 
characteristic velocity dispersions (corresponding to the characteristic 
absolute magnitudes) were then constrained by the observed relative
frequencies of image multiplicities and the observed image 
separations in the final CLASS statistical sample. We found that
the values of the early-type characteristic velocity dispersion obtained in
the way described above agreed well with the values based on the early-type
characteristic absolute magnitude and a Faber-Jackson relation or a 
$\sigma_{\rm DM}$-$L$ relation (recall here that a singular isothermal
mass profile was also assumed; see section~4).  
The values of the late-type characteristic velocity dispersion
obtained from the statistical lensing analyses were $\approx 1\sigma$ lower 
than the value based on the late-type characteristic absolute magnitude and a 
Tully-Fisher relation. 

How would the derived results from the statistical lensing analyses be
affected if non-isothermal mass profiles, non-axisymmetric intrinsic shapes,
and/or general three-integral distribution functions of galaxies were 
considered? Constraints on cosmological parameters are not likely to be 
affected by non-isothermal mass profiles, non-axisymmetric intrinsic shapes,
or general distribution functions according to the following arguments.
For the given apparent axial ratio and line-of-sight velocity dispersion,
the cross sections averaged over the allowed ensemble of models obviously
depend on the assumed galaxy model. 
Hence, the derived characteristic velocity dispersions 
from the observed image separations depend on the mass profile,
the allowed range of intrinsic shapes, and the assumption on the distribution
function of the galaxy. (Dependences on the intrinsic shapes for the
singular isothermal mass profile were illustrated in section~4.) However,
while the derived characteristic velocity dispersions are affected as
the galaxy model is varied, the multiple-imaging cross sections are not
as long as the projected surface density of the inner cylindrical region of
the galaxy is similar to the isothermal surface density. 
In other words, the observed image separations will determine the 
multiple-imaging cross sections regardless of the details
of the galaxy model as long as the image separation 
for the given multiple-imaging cross section is not significantly 
dependent on the image magnification ratio, similarly to the case
for the singular isothermal profile in which the image separation is
fixed by the cross section and independent of the magnification ratio.

\subsubsection{Local early-type LF: the characteristic number density
and faint-end slope}

As discussed in section~3.3, there is no consensus
among observationally derived results for the early-type LF.
We have used two choices of the early-type LF in this work, namely
the steep SSRS2 choice and the shallow 2dFGRS choice.
As seen in section~4, while the two different early-type LFs require
different values of $\sigma_{*}^{(e)}$ to fit the same image separations
in the CLASS statistical sample, the difference in the
values of $n_{*,0}^{(e)}$ for the two early-type LFs is such that 
the two early-type LFs give similar results on cosmological parameters
({Fig.}~6 \& 7). However, the fact that the two different 
early-type LFs give similar results on cosmological parameters
does not preclude the possibility that the presently derived constraints on
cosmological parameters based on the two early-type LFs suffer from
systematic errors due to the present uncertainties in the early-type LF.
As pointed out in section~3.3, the main difficulty in the observational 
derivation of the early-type LF lies in classifying large numbers of galaxies
by morphological types. Systematically misclassifying galaxies will
not only result in an error in the number density of early-type galaxies
but also an error in the faint-end slope in the early-type LF. 
Given that the derived cosmological parameters depend on both the early-type
characteristic number density and faint-end slope, it is understood that
the presently derived constraints on cosmological parameters are potentially
susceptible to unquantified systematic errors due to the present uncertainties
in the early-type LF. Nonetheless, since the total galaxy LF has been reliably
determined, as long as the partition of the total LF into the type-specific
LFs is not drastically different from those considered in this work, the 
conclusions drawn on cosmological parameters (section~4.1) would not be 
significantly affected.

\subsubsection{Evolution of early-type galaxies}
One key assumption made in this work is no evolution of early-type
galaxies since $z \sim 1$ in the sense that the early-type characteristic
comoving number density and faint-end slope 
are unchanged from $z \sim 1$ to the present epoch.
In section~3.3, we pointed out that the hypothesis of the early ($z \ga 2$)
formation and passive evolution of early-type galaxies is supported by
many lines of observational studies.

However, the evolution of early-type galaxies remains a relatively 
larger source of uncertainty for analyses of statistical lensing.
First, the observational results that provided the most direct pieces of 
evidence for little or no evolution of the early-type characteristic
comoving number density and faint-end slope were based on relatively
small samples of galaxies and were less than abundant. Furthermore,
limits on galaxy evolution derived from galaxy counts are dependent on
the assumed cosmological model and thus, in a strict sense, should not
be used for constraining cosmological parameters via statistical lensing
(see, e.g., Keeton 2002).
Second, there exist observational results in the literature that have been 
used to claim a rapid evolution of early-type galaxies in which there is a 
significant reduction in the comoving number density of early-type galaxies
at intermediate redshifts ($0.3 \la z \la  1$) compared with the present
epoch.\footnote{However, it is interesting to note that those who
found a rapid evolution of early-type galaxies classified galaxies through
photometric/spectroscopic information (e.g.\ Fried et al.\ 2001)
while those who found little or no evolution of early-type galaxies 
classified galaxies through morphological appearances or light profiles
(e.g.\ Im et al.\ 2002).}

If future observational studies established a rapid evolution in the 
population of early-type galaxies since $z \sim 1$ 
so that the comoving number density of early-type galaxies
at an intermediate redshift was significantly lower than that at the present
epoch, the constraints on cosmological parameters derived in this work would
have to be adjusted; the adjustment would increase (decrease) dark energy 
density (matter density). Reversing the problem, statistical lensing can
be used in the future to constrain the formation and evolution of early-type
galaxies assuming a certain cosmological model. {From} the results of 
this work, the qualitative inference on the formation and evolution of 
early-type galaxies would be that a flat `concordance cosmology' with
$\Omega_{\rm m} \approx 0.3$ ($\Omega_{\Lambda} \approx 0.7$) favors an early
($z \ga 1.5$) formation of early-type galaxies followed by passive evolution.
A detailed quantitative analysis of statistical lensing 
focusing on constraining the formation and evolution of
early-type galaxies assuming the concordance cosmology will be presented in
a forthcoming publication (Mao \& Chae 2003).

\subsubsection{Redshift distribution of CLASS sources}
As shown in Figure~10(a), the derived value of $\Omega_{\rm m}$ (in a flat
universe) is sensitive to the mean redshift of CLASS sources. 
Since the Marlow et al.\ (2000) measurement has a relatively large 
uncertainty due to the relatively small number of 
measured spectroscopic redshifts, it will be important to obtain
secure redshifts for more CLASS sources in the future to reduce the 
uncertainty arising from the current uncertainty in the mean redshift
of CLASS sources.

Six multiply-imaged sources in the final CLASS statistical sample (Table~1)
lack measured redshifts. While the unknown redshifts would not change the
derived results from analyses of statistical lensing as long as they are not 
too close to their corresponding lens redshifts,\footnote{The uncertain 
redshift of 2045+265 can be an issue here. The system has the highest lens 
redshift among the measured lens redshifts and a relatively large image 
separation.
Furthermore, there was an early report that the source redshift could be
$z_s = 1.28$ although this interpretation was very uncertain 
(Fassnacht et al.\ 1999).} it would be important to measure them not just 
as a matter of completeness but to make sure that no surprising results 
are missed. 

\section{Future prospects of statistical lensing}
As demonstrated in detail in this work, statistical properties of strong
gravitational lensing allow us to probe independently cosmological models and
global properties of galaxies. In principle, a sample of cosmologically
distant sources that is observationally well-defined and large enough can
alone pin down both cosmological parameters (e.g.\ $\Omega_{\rm m}$, 
$\Omega_{\Lambda}$, and $\omega$) and global parameters of galaxies
(e.g.\ $\sigma_*^{(e)}$, early-type velocity function slope,
$\sigma_*^{(s)}$, spiral-type velocity function slope, and
$\bar{\epsilon}_{\rm mass}$) including their evolutionary properties. 
Since determining cosmological parameters and global properties of galaxies 
is a main and fundamental part of cosmology, it is important to have various
independent methods. Statistical lensing provides such a method. 

The Cosmic Lens All-Sky Survey is the largest systematic strong gravitational
lens survey at the time of this writing. A second Cosmic Lens All-Sky Survey
(CLASS2) is being planned and would aim to discover ten times 
as many multiply-imaged radio-loud sources as the CLASS did.
A statistical sample from a CLASS2 project would allow us to measure
cosmological and galactic parameters with three times smaller statistical
errors compared with the values obtained in this work assuming that the
errors are Poissonian [this appears to be the case from the comparison of 
{Fig.}~12~(c) or (d) with {Fig.}~6~(a) (see section~5.2) and we will deal 
with the issue regarding errors in a future publication through an extensive
Monte Carlo simulation of statistical lensing (Chae et al., in preparation)].
For example, this means that statistical lensing could independently come up
with a value of $\Omega_{\rm m}$ (or $\Omega_x$) with an error of $\sim 0.05$
(if a flat universe is assumed) and values of 
$\sigma_*^{(e)}$ and $\sigma_*^{(s)}$ with errors of $\sim 7$~km~s$^{-1}$.

It is almost certain that statistical lensing will not enable us to measure 
cosmological parameters with the precision that is achievable by cosmic
microwave background anisotropy measurements or supernovae observations
through space-based instruments for the present and the coming decade [e.g.,
Wilkinson microwave anisotropy probe ({\it WMAP}: Spergel et al.\ 2003); 
Planck surveyor\footnote{http://astro.esa.int/SA-general/Projects/Planck};
supernova/acceleration probe (SNAP)\footnote{http://snap.lbl.gov}]. 
Nevertheless, independent results on cosmological parameters from statistical
lensing are of interest and value because the importance of cosmological
parameters warrants demanding tests. However, perhaps the greatest value of
statistical lensing for the coming years lies in its unique probe of the 
velocity functions of galaxies and galaxy evolution. In particular, the 
velocity dispersions implied by the image separations of strongly-lensed 
systems are different from those measured spectroscopically from galaxy
surveys such as the SDSS, and statistical lensing probes galaxy evolution
based on mass-selected samples of galaxies whereas studies from galaxy
surveys are based on brightness-selected samples.
Thus, statistical lensing provides valuable and unique constraints on
theories of the formation and evolution of galaxies.

\section{CONCLUSIONS}
{From} the analyses of statistical lensing based on the final CLASS statistical
sample, the current estimate of luminosity functions of galaxies per 
morphological type, and the singular isothermal ellipsoid (SIE) lens model, 
we have reached the following conclusions:
\begin{enumerate}
\item The multiple-imaging cross sections of galaxies implied by the observed
image separations are such that the observed `local' distributions of galaxies
require a cosmological model dominated by dark energy to be compatible with
the observed multiple-imaging rate, under the assumption that field 
early-type (i.e.\ E/S0) galaxies existed with the present-epoch abundance 
and masses at intermediate redshifts $0.3 \la z \la 1$. In particular,
the confidence region in the $\Omega_{\rm m}$-$\Omega_{\Lambda}$ plane is 
in good agreement with that from Type~Ia supernovae observations
(Riess et al.\ 1998; Perlmutter et al.\ 1999). 
This independent result further supports a flat concordance cosmology 
with $\Omega_{\rm m} \sim 0.3$ ($\Omega_{\Lambda} \sim 0.7$). However,
if the early-type galaxy population underwent a significant evolution from 
$z \sim 1$ to the present epoch as in a certain hierarchical galaxy formation
picture, the statistical lensing data used in this work would imply an even
higher dark energy density.

\item Assuming that one half of early-type  galaxies are oblates 
and the other half are prolates, the fitted early-type characteristic 
line-of-sight velocity dispersion is
$\sigma_{*}^{(e)} (0.3 \la z \la 1) = 198^{+22}_{-18}$~km~s$^{-1}$ (68\%) 
if the early-type LF faint-end slope  follows that of the `steep'
SSRS2 early-type LF (i.e.\ $\alpha^{(e)}=-1$), or 
$\sigma_{*}^{(e)} (0.3 \la z \la 1) = 181^{+18}_{-15}$~km~s$^{-1}$ (68\%) 
if it follows that of the `shallow' 2dFGRS early-type LF 
(i.e.\ $\alpha^{(e)}=-0.54$). 
These results are consistent with the current estimate of the early-type
characteristic absolute magnitude in the $B$ band and the $\sigma$-$L$
relation, but are significantly lower than the previous result based on
the optically-selected lens sample and the SSRS2-like steep early-type 
faint-end slope. 

\item Assuming that one half of early-type galaxies are oblates and the 
other half are prolates, the best-fit value for the mean projected mass 
ellipticity of early-type galaxies is 0.42 and the 68\% lower limit is 0.28.
The inferred mean projected mass ellipticity agrees well with the measured 
mean light ellipticity of early-type galaxies in `fields' (i.e., random
environments) or in clusters. 

\end{enumerate}

\section*{Acknowledgments}
This work is the follow-up of, and complementary to, Chae et al.\ (2002), 
in which the CLASS team reports constraints on cosmological parameters 
based on the final CLASS data. As a `latest member' of the CLASS 
collaboration, the author thanks those who worked hard to produce the final 
CLASS data. J. McKean et al.\ are further thanked for their work on
CLASS parent source counts.
The author particularly thanks Ian Browne and Peter Wilkinson for
initial motivational comments that led him to contribute toward the Chae 
et al.\ (2002) results and this work. Ian Browne has also provided useful
comments on the text. The author thanks the anonymous
referee of the paper for a thorough review and numerous 
constructive comments. In particular, we owe the referee {Fig.}~12 and relavant 
discussion. The additional work presented in the revised version was supported
by the Astrophysical Research Center for the Structure and Evolution of the 
Cosmos (ARCSEC) which was established  under the Science Research Center 
(SRC) program of Korea Science and Engineering Foundation (KOSEF).

{}

\bsp
\label{lastpage}
\end{document}